\documentclass[prd,a4paper,nofootinbib,notitlepage,twocolumn,showpacs]{revtex4-1}

\usepackage[utf8x]{inputenc}
\usepackage{amssymb}
\usepackage{amsmath}
\usepackage{graphicx}
\usepackage{psfrag}
\usepackage{latexsym}
\usepackage{hyperref}
\usepackage{longtable}

\usepackage{pbox}

\newcommand{\eq}{\begin{equation}}
\newcommand{\eqx}{\end{equation}}
\newcommand{\eqn}{\begin{eqnarray}}
\newcommand{\eqnx}{\end{eqnarray}}

\newcommand{\f}[2]{\frac{#1}{#2}}

\newcommand{\nn}{{\mathcal N}}
\newcommand{\LL}{{\mathcal L}}

\newcommand{\eps}{\varepsilon}

\newcommand{\Lm}{\Lambda}
\newcommand{\gm}{\gamma}
\newcommand{\al}{\alpha}
\renewcommand{\alt}{\tilde{\alpha}}
\newcommand{\alphat}{\tilde{\alpha}}

\newcommand{\du}{\partial_u}
\newcommand{\dt}{\partial_t}

\newcommand{\oo}[1]{{\cal O}\left(#1\right)}

\begin{document}

\title{A numerical relativity approach to the initial value problem in
asymptotically Anti-de Sitter spacetime for plasma thermalization --- an ADM formulation}

\author{Michal P. Heller}\email{m.p.heller@uva.nl}
\altaffiliation[On leave from: ]{\it National Centre for Nuclear Research,
  Ho{\.z}a 69, 00-681 Warsaw, Poland}
\affiliation{\it Instituut voor Theoretische Fysica, Universiteit van Amsterdam, Science Park 904, 1090 GL Amsterdam, The Netherlands}

\author{Romuald A. Janik}\email{romuald@th.if.uj.edu.pl}

\author{Przemys{\l}aw Witaszczyk}\email{bofh@th.if.uj.edu.pl}

\affiliation{Institute of Physics,
Jagiellonian University, Reymonta 4, 30-059 Krak\'ow, Poland}

\begin{abstract}
This article studies a numerical relativity approach to the initial value problem in Anti-de Sitter spacetime relevant for dual non-equilibrium evolution of strongly coupled non-Abelian plasma undergoing Bjorken expansion. In order to use initial conditions for the metric obtained in \href{http://arxiv.org/abs/arXiv:0906.4423}{\tt arXiv:0906.4423}  we introduce new, ADM formalism-based scheme for numerical integration of Einstein's equations with negative cosmological constant.  The key novel element of this approach is the choice of lapse function vanishing at fixed radial position, enabling, if needed, efficient horizon excision. Various physical
aspects of the gauge theory thermalization process in this setup have been outlined in our companion article \href{http://arxiv.org/abs/arXiv:1103.3452}{\tt arXiv:1103.3452}. In this work we focus on the gravitational side of the problem and present full technical details of our setup. 
We discuss in particular the ADM formalism, the explicit form of initial states, 
the boundary conditions for the metric on the inner and outer edges of the simulation domain, the relation between boundary and 
bulk notions of time, the procedure to extract the gauge theory energy-momentum 
tensor and non-equilibrium apparent horizon entropy, as well as the choice of 
point for freezing the lapse. Finally, we comment on various features of the initial profiles we consider.
\end{abstract}


\maketitle

\twocolumngrid

\section{Introduction}

The most fascinating theoretical challenge in the physics of heavy ion 
collisions is the understanding of the mechanisms behind the short -- less 
than $1$ fm/c \cite{Heinz:2004pj} time after which a hydrodynamic description is 
neccessary in order to describe experimental data. Due to the fact that
hydrodynamics typically assumes local thermal equilibrium, this problem has
been dubbed `the early thermalization problem'. 
A possibility which has not been fully appreciated, is that viscous hydrodynamic
description may be applicable even with quite sizable pressure anisotropy \cite{Chesler:2008hg,Heller:2011ju,Strickland:2011}. Thus
the quark-gluon plasma may still not be in a real thermalized state even though
all the experimentally observed consequences of a (viscous) hydrodynamic description
may still hold. 

Because of this experimental motivation, and the adopted usage in the heavy-ion
community, we will continue to call the transition to viscous hydrodynamics as 
`thermalization' even if the plasma is not really strictly thermalized in the
statistical mechanics sense. The question, to what extent is the plasma
isotropic at this `thermalization' is at the forefront of the current investigations.

The very low observed shear viscosity of the plasma
produced in relativistic heavy-ion collision strongly suggests that the
plasma system is strongly coupled, thus making the theoretical analysis of the 
thermalization problem very difficult if not impossible\footnote{In QCD we may
even have a mixture of perturbative and non-perturbative effects.}.

On the other hand, this observation opens up the possibility of using
methods of the AdS/CFT correspondence to study similar problems in gauge theories
which possess a gravity dual. The utility of the AdS/CFT correspondence in the
non-perturbative regime lies in the fact that then the lightest dual degrees of 
freedom at strong coupling are just the (super)gravity modes whose dynamics is
given by Einstein's equations (possibly with specific matter fields).
Moreover, from the metric sector one can extract the whole dynamics of the
expectation values of the gauge theory energy-momentum tensor which is the
key observable of interest in all hydrodynamic models.

Thus Einstein's equations (more precisely supergravity equations) are an ideal
arena to study the physics of thermalization in a strongly coupled gauge theory.
This setup takes the simplest form in $\nn=4$ super Yang-Mills theory (SYM),
when we do not excite any other expectation values apart from the energy-momentum
tensor. Then the gravitational description reduces to 5-dimensional Einstein's
equations with negative cosmological constant. Even more interestingly,
this description is universal in this sector for all holographic 1+3 dimensional
conformal field theories as argued in \cite{Bhattacharyya:2008mz}.

A static system in thermal equilibrium is decribed on the dual side by a planar
black hole. It is known since \cite{Horowitz:2002ab} that thermalization of
generic small disturbances of the system is exponentially fast and is described,
on the gravitational side, by quasi-normal modes.

For an expanding plasma system (e.g. in the boost-invariant case considered here), 
the asymptotic equilibrium geometry is time dependent and looks like a boosted
black hole and the thermalization of small disturbances is also very fast but now 
quasi-exponential (i.e. $\sim \exp(-const\, \tau^{\f{2}{3}})$) \cite{Janik:2006ax}.
In that paper, thermalization was suggested to occur as an approach to
an attractor geometry from generic initial conditions, but clearly a linearized
analysis is insufficient\footnote{But c.f. the recent work \cite{Heller:2012km} 
for an in-depth analysis showing the effectiveness of using linearized methods 
in the case of isotropization.} -- a full nonlinear treatment of Einstein's 
equations
is needed thus neccesitating a numerical approach.

Intuitively, the most important feature of gravity backgrounds dual to collective states of matter is the presence of the (event) horizon, which acts as a membrane absorbing gravitational radiation and matter outside until local equilibrium (in the sense of perfect fluid hydrodynamic description) is reached on the boundary. This mechanism of equilibration turns out to be very effective. Indeed numerical simulations of
\cite{Chesler:2008hg,Chesler:2009cy,Chesler:2010bi} give short thermalization 
times (which can be argued to correspond to times shorter than $1$ fm/c
at RHIC energies). In addition, the result of our 
investigation~\cite{Heller:2011ju} shows that viscous hydrodynamics applies 
for $T\tau_{th}\geq 0.6-0.7$ which is consistent
with RHIC assumptions (e.g. $T=500MeV$ and $\tau_{th}=0.25fm$ gives $T\tau=0.63$).
These results provide very strong motivation for further investigations of the 
thermalization processes in the gauge-gravity duality\footnote{We do not discuss
here the very interesting investigations of thermalization in $\nn=4$ SYM on 
$S^3\times R$~\cite{Bizon:2011ax,Horowitz:2011bx,Garfinkle:2011av}.}.

What makes the thermalization process difficult are the many scales involved, 
so that the full microscopic description is needed. This is in stark contrast 
with the near-equilibrium dynamics, where the evolution of the system is governed 
by the equations of hydrodynamics and the only microscopic input needed 
are thermodynamic relations and lowest transport coefficients (up to first or 
second order in gradients). A beautiful holographic manifestation of this fact 
is the fluid-gravity duality \cite{Bhattacharyya:2008jc}, where the velocity 
and temperature profiles on the boundary specify completely the dual gravity 
background and Einstein's equations can be recast in the form of the equations 
of hydrodynamics. In other words, once holography provides the thermodynamic 
and transport properties of the dual field theory, in order to solve the initial 
value problem in the near-equilibrium regime, one does not need to solve the 
full Einstein's equations. The same information is contained in the much easier 
equations of hydrodynamics.

On the other hand, in the far-from-equilibrium regime in a holographic field theory, 
one needs to specify not a couple (as in hydrodynamics), but an infinite number 
of boundary functions (a couple of numbers at \emph{each} constant radius slice 
of the initial time hypersurface in the bulk). Gravity then is just a clever 
way of recasting complicated interactions between these degrees of freedom 
in the form of classical equations of motion for the five-dimensional metric. 

From that perspective, the study of far-from-equilibrium physics leading to 
thermalization consists of two natural steps. First, one needs to specify the 
initial data for the evolution on some initial bulk time hypersurface and later 
use fully-fledged numerical relativity to evolve it till the dual stress tensor 
can be described by hydrodynamics. The thermalization time is then the boundary 
time which elapsed between these two events.

Because the thermalization process is so complicated, one of the most important questions is whether any regularities emerge from underlying microscopic (here gravitational) dynamics. Another way of looking at this is to search for some characteristic features of holographic, and so driven by strong coupling, thermalization, which might provide clues for singling out mechanisms relevant for a rapid approach to local equilibrium at RHIC. Yet another perspective is to try to understand various quantities describing thermalization at strong coupling in terms of some primary and secondary features of initial far-from-equilibrium state encoded geometrically. 

Motivated by these questions we are considering a boost-invariant \cite{Bjorken:1982qr} thermalization process in the holographic conformal setting. The reason for focusing on the boost-invariant flow is that it is phenomenologically relevant for the mid-rapidity region of heavy ion collisions, but at the same time simple enough to allow for a thorough understanding using the gauge-gravity duality. In the boost-invariant case with no transverse dynamics the evolution of finite energy density system depends on a single coordinate -- proper time $\tau$ -- and necessary starts with some far-from-equilibrium state at early time, which thermalizes at some transient time leaving viscous hydrodynamics at late time.

The holographic studies of the boost-invariant flow have a long history and started with unravelling perfect \cite{Janik:2005zt} and further first \cite{Janik:2006ft}, second \cite{Heller:2007qt} and third order \cite{Booth:2009ct} hydrodynamics in the late time expansion of dual spacetime, subsequently embedded within the framework of the fluid-gravity duality \cite{Bhattacharyya:2008jc} as its special case \cite{Heller:2008mb,Kinoshita:2008dq,Kinoshita:2009dx}.

The key later development underlying the present article is \cite{Beuf:2009cx}, where a Taylor expansion in the Fefferman-Graham (FG) coordinates was used to unravel the structure of 
the metric coefficients at early time and the nature of dual far-from-equilibrium initial state. Complementary to the studies of boost-invariant holographic thermalization reported in \cite{Chesler:2009cy} this does not require turning on sources for any field theory operators and allows to observe unforced relaxation towards local equilibrium starting from a family of far-from-equilibrium initial states. The crucial simplification occurring in the Fefferman-Graham coordinates when the initial time hypersurface in the bulk is chosen to be $\tau_{FG} = 0$, is that the regularity of bulk geometry forces time derivatives of all 3 nontrivial metric coefficients 
(warp-factors) to vanish. This allows to solve explicitly 2 constraint equations and parametrize the most general initial data in terms of a single function of radial direction subject to regularity constraint and asymptotic AdS boundary condition. 

The coefficients of near-boundary expansion of the warp-factor specifying initial data turns out to be related to the early time expansion of the stress tensor in a one-to-one fashion, in accord with expectation that far-from-equilibrium dynamics involves many independent scales. By reconstructing early time power series from the near-boundary behavior of initial data one could not however see the transition to hydrodynamics, as the resulting expressions have too small radius of convergence in $\tau$, which directly motivates this work.

The task of this article is to provide a numerical framework which allows to solve Einstein's equations starting from initial data found in \cite{Beuf:2009cx} and evolve them till hydrodynamic regime on the boundary is reached. The most important 
physical results about the thermalization process in this setting were summarized in the letter \cite{Heller:2011ju}, whereas this companion article focuses on the gravity side of our approach 
and its numerical formulation.

The principal complication in formulating a numerical framework is the well known
diffeomorphism-invariance of General Relativity. The form of the equations
depend on the type of coordinate system that one uses, and in addition, depend
on the choice of dynamical variables. A good choice should be well adapted to the
physical problem under investigation and of course should be stable numerically.

One edge of the computational domain can be naturally taken as the boundary of asymptotically AdS spacetime with boundary conditions dictated by imposing the flatness of the metric in which the dual field theory lives. This condition is imposed because we do not want to deform the gauge theory through giving
a source to its stress tensor or any other operator, but rather study unforced 
relaxation from a set of far-from-equilibrium initial data.

Typically one has also to impose boundary conditions at the other (outer) edge of 
the integration domain, where generically the curvature may be quite high.
In a standard formulation, causality of the bulk spacetime requires putting this edge behind the event horizon, so that initial data specified on an initial time hypersurface encode boundary dynamics up to an arbitrary late time. However, the event horizon is a global concept and cannot be located on a given slice of constant time foliation until the full spacetime is known. 
As a measure whether a given point is outside or inside the event horizon one can use the notion of an apparent horizon, which is intrinsic to a given slice. On the other hand, the existence and the location of an apparent horizon depends on a foliation and a constant time foliation induced by a given choice of coordinate frame might not see (at least initially) any apparent horizon.

The most commonly used method of integrating Einstein's equations in asymptotically
AdS spacetimes was to use a characteristic formulation and use ingoing 
Eddington-Finkelstein (EF) coordinates \cite{Chesler:2008hg,Chesler:2009cy,Chesler:2010bi}.  
This method has been very successful in numerical simulations of black hole 
formation in AdS spacetimes. Here, the outer boundary can be set to be at the
location of the apparent horizon on the null hypersurface. Due to the null
character no explicit boundary conditions need to be set there.

For our purposes we did not adopt this formulation as the ingoing null light rays, do not align along $\tau_{FG} = 0$ hypersurface and so initial conditions derived in \cite{Beuf:2009cx} and reviewed in section \ref{sec.bif} cannot be used as a starting point for the Eddington-Finkelstein approach. There seems to be also an unresolved technical problem in writing numerical codes in Eddington-Finkelstein coordinates starting from $\tau = 0$ at the boundary stemming from the fact that in these coordinates the limits $\tau_{EF} \rightarrow 0$ and $r \rightarrow \infty$ (going to the boundary) do not commute (see section \ref{sec.bif} for more details). 

These difficulties forced us to search for a new coordinates frame in which the hypersurface $\tau_{new} = 0$ coincides with the hypersurface $\tau_{FG} = 0$ and which provides a sensible radial cutoff in the bulk enabling numerical treatment. Fortunately, it turned out that successful code can be achieved by adopting to AdS gravity 
the ADM (Arnowitt-Deser-Misner) formalism of general relativity \cite{MWT-ADM}
with fixed time hypersurfaces being spacelike. 
This scheme is also the most popular in numerical simulations in asymptotically flat spacetimes, but so far, to our knowledge, has not been used in the AdS/CFT context. Its advantage is also that it is a very generic formulation of
an initial value problem, thus it should be easy to generalize to other setups.

The ADM formulation for an asymptotically AdS spacetime involved two difficulties
which were not present in conventional asymptotically flat formulations. Firstly,
the boundary conditions at the AdS boundary, dictated e.g. by the choice that 
the gauge theory metric is Minkowski, turned out to be surprisingly subtle and 
involved a careful treatment of a possible boundary diffeomorphism.
Secondly, and this is the chief novel feature of our formulation, we adopted
the outer boundary conditions by freezing the evolution at the outer edge
by making the lapse vanish there. This construction has several attractive
features. It works perfectly well even when the geometry is highly curved at the 
edge. The exterior of the simulation domain is causally disconnected from
the interior and thus the obtained results are completely determined by the
initial data. This last feature is \emph{not} dependent on the location
of the edge w.r.t. the event horizon. Thus we may perform numerically consistent
simulations without any knowledge on the location of the event horizon.

Let us finally note, that very recently a third type of numerical relativity
formulation was adapted to Asymptotically AdS spacetimes --- the Generalized
Harmonic formulation \cite{Pretorius:2012zx}. It would be very interesting to 
investigate its relative merits with the ADM formulation (in particular in its 
more refined versions like BSSN ~\cite{Shibata:2009ax}) in the present context.

The plan of the article is the following. Section \ref{sec.bif} introduces the boost-invariant flow in the context of the gauge-gravity duality and reviews the results of \cite{Beuf:2009cx}. Section \ref{sec.newcoord} explains how to bypass the Fefferman-Graham coordinate frame singularity by a choice of chart inspired by the Kruskal-Szekeres extension of Schwarzschild metric. Section \ref{sec.admreview} reviews ADM formalism of general relativity, which in section \ref{sec.admforbif} is tailored to describe the gravity dual to boost-invariant flow. Section \ref{sec.observables} explains the subtleties of imposing asymptotically AdS boundary conditions and obtaining expectation value of dual stress tensor operator, as well as elaborates on non-equilibrium entropy defined on the apparent horizon. Section \ref{sec.numsim} describes various aspects of numerical side of the project. Section \ref{sec.profiles} elaborates on the analyzed initial conditions, compares numerical predictions for the effective temperature as a function of proper time with early time power series for three representative initial profiles, as well as explains subtlety in defining thermalization time for a class of initial conditions we considered. Finally, the last section summarizes results and discusses open problems.

\section{Boost-invariant flow and holography \label{sec.bif}}
\subsection{Kinematics}

Boost-invariant dynamics describes an expansion of plasma with an additional 
assumption that physics remains the same in all reference frames boosted along 
the expansion axis (i.e. in the longitudinal plane). This symmetry can be made 
manifest by introducing proper time $\tau$ and (spacetime-)rapidity 
$y$ coordinates related to the usual lab frame time $x^{0}$ and position along 
the expansion axis $x^{1}$ by
\eq
\label{bicoord}
x^{0} = \tau \, \cosh y \quad \mathrm{and} \quad x^{1} = \tau \, \sinh y.
\eqx
In the following $x^{2,3}$ are taken to be Cartesian coordinates in the transverse plane and are denoted collectively as $x_{\perp}$. In the absence of transverse dynamics, which is the simplifying assumption adopted here, the evolution of the system in proper time -- rapidity coordinates depends only on proper time, since boosts along $x^{1}$ direction shift rapidity. The background Minkowski metric in 
these coordinates becomes proper time-dependent
\eq
\label{mink.metric.in.bi.coord}
d s_{boundary}^{2} = \eta_{\mu \nu} dx^{\mu} dx^{\nu} = - d\tau^{2} + \tau^{2} dy^{2} + dx_{\perp}^{2}
\eqx
so that, as anticipated in the introduction, the system, by construction, is not translationally invariant in proper time and its evolution naturally splits into the early, transient and late time dynamics even if no external work is done on the system. It is also worth noting at this point that proper time - rapidity coordinates are curvilinear and despite the fact that the boost-invariant dynamics depends only on a single timelike coordinate, there is a hydrodynamic tail at late time in contrast to spatially uniform isotropization \cite{Chesler:2008hg,Heller:2012km}.

The field theory observable, which is of interest here, is the expectation value 
of the energy-momentum tensor operator. This object carries direct information whether the system is in local equilibrium and, in holographic conformal field theories, undergoes decoupled dynamics specifying by itself dual gravity background \cite{Bhattacharyya:2008mz}. The most general traceless and conserved stress tensor obeying the symmetries of the problem in the coordinates \eqref{bicoord} takes the form
\eq
\label{gen.stress.tensor}
T^{\mu}_{\,\,\, \nu} = diag \left( - \eps, p_{L}, p_{T}, p_{T} \right),
\eqx
where $p_{L}$ and $p_{T}$ are longitudinal and transverse pressures expressed in terms of energy density $\eps(\tau)$ and read
\eq
\label{plANDpt}
p_{L} =  - \eps - \tau \, \eps' \quad \mathrm{and} \quad p_{T} =  \eps + \frac{1}{2} \tau \, \eps'.
\eqx
The precise form of the energy density as a function of time $\eps(\tau)$ 
depends on the initial state and is governed by the complicated dynamics of a gauge theory or, here, by the dual gravity picture.
The principal aim of the present investigation is to devise a method to
obtain $\eps(\tau)$ for any given initial conditions.

\subsection{Bjorken hydrodynamics and the criterium for thermalization\label{subsec.bjorkenhydro}}

At sufficiently late times boost-invariant plasma evolves according to the equations of hydrodynamics \cite{Bjorken:1982qr}. These are conservation equations of the stress tensor \eqref{gen.stress.tensor} under the
assumption that it can be written in hydrodynamic form, i.e. expressed in terms of a local temperature, velocity and gradients of velocity. 
Let us note that this assumption is not true in general. Whether one can
write $T_{\mu\nu}$ in such a form is a question of dynamics. Below, following \cite{Heller:2011ju} we
will adopt an unambiguous criterion testing whether in the boost-invariant setup $T_{\mu\nu}$ can indeed be written in 
hydrodynamic form or not.

As symmetries -- boost-invariance, invariance under reflections in rapidity, as well as rotational and translational symmetries in the transverse plane -- fix the form of the local velocity (its only non-zero component is $u^{\tau} = 1$), the only non-trivial dynamical hydrodynamic field in the setup of interest is the local (called here effective for the reasons explained below) temperature $T_{eff}(\tau)$ defined by
\eq
\label{def.effective.temp}
\eps(\tau) = \frac{3}{8} N_{c}^{2} \pi^{2} T_{eff}(\tau)^{4}.
\eqx
The scaling of the energy density \eqref{def.effective.temp} with local temperature is fixed by the scale invariance of gauge theory of interest, whereas the prefactor counting the number of degrees of freedom in thermodynamic equilibrium is that of $\mathcal{N} = 4$ super Yang-Mills at large $N_c$ and strong coupling.

We will use the equation (\ref{def.effective.temp}) also in the far-from-equilibrium regime, but there we will treat it as a {\it definition}
of an effective temperature. Physically this is the temperature of
a thermal state of $\nn=4$ SYM theory with the same energy density
as $\eps(\tau)$. This gives us a measure of the energy density which factors out the number of degrees of freedom relevant for the specific gauge theory.
In the rest of the text we will refer to $T_{eff}(\tau)$ as to an effective temperature.

It is easy to see, that the equations of hydrodynamics in the boost-invariant setup reduce to a first order ordinary differential equation for the effective temperature, which can be solved perturbatively in the large proper time expansion. The result is known explicitly up to the third order in gradients for $\mathcal{N}=4$ super Yang-Mills and other $(3+1)$-dimensional holographic conformal field theories \cite{Janik:2006ft,Heller:2007qt,Booth:2009ct} reading
\begin{eqnarray}
\label{T.hydro}
T_{eff}(\tau) =&& \frac{\Lm}{\left( \Lm \tau \right)^{1/3}} \Big\{ 1 - \frac{1}{6 \pi \left( \Lm \tau \right)^{2/3}} + 
\frac{-1 + \log{2}}{36 \pi^{2} \left( \Lm \tau \right)^{4/3}} + \nonumber\\
&&+ \frac{-21 + 2\pi^{2} + 51 \log{2} - 24 \log{2}^{2}}{1944 \pi^{3} \left(\Lm \tau\right)^{2}}+ \ldots
\Big\},
\end{eqnarray}
where $\Lm$ is an integration constant with a dimension of energy governing the asymptotic scaling of the effective temperature with proper time \cite{Chesler:2009cy,Heller:2011ju}. Given the effective temperature as a function of time, $\Lm$ can obtained by fitting the hydrodynamic expression for $T_{eff}(\tau)$ to late time data obtained from numerical simulation.


Introducing a dimensionless variable $w$ being the product of the effective temperature and proper time
\eq
\label{wdef}
w(\tau) \equiv T_{eff}(\tau) \tau
\eqx
allows to rewrite the equation of boost-invariant hydrodynamics in a particularly simple form reading
\eq
\label{eq.Fw}
\f{\tau}{w} \f{d}{d\tau} w = \f{F_{hydro}(w)}{w},
\eqx
where $F_{hydro}(w)$ is a universal function of $w$. Hydrodynamic gradient expansion coincides with the large-$w$ expansion of $F_{hydro}(w)$. 
This approach to boost-invariant hydrodynamics is similar in spirit to previous attempts of Shuryak and Lublinsky of introducing all-order resummed hydrodynamics \cite{Lublinsky:2007mm,Lublinsky:2009kv}.
The important difference is that $F_{hydro}(w)$ contains \emph{all}
nonlinear hydrodynamic effects. Using the results of \cite{Janik:2006ft,Heller:2007qt,Booth:2009ct} we calculated in \cite{Heller:2011ju} $F_{hydro}(w)$ up to third order in gradients obtaining
\eqn
\label{e.third}
&& \f{F_{hydro}(w)}{w} = \f{2}{3}+ \f{1}{9\pi w} +
\f{1-\log 2}{27\pi^2 w^2}+ \nonumber \\
&& +\f{15-2\pi^2-45\log 2+24 \log^2 2}{972 \pi^3 w^3} +\ldots
\eqnx
Let us note that even without knowing the precise form of $F_{hydro}$,
equation \eqref{eq.Fw} can be used as a test whether $T_{\mu\nu}$ is in
a hydrodynamic form (i.e. written purely in terms of a local temperature and
gradients of velocity). Indeed, plotting the left hand side of (\ref{eq.Fw})
as a function of $w$ for various initial conditions would give a 
\emph{single} curve if the hydrodynamic description would be valid or 
\emph{multiple} curves in the opposite case. This analysis was performed
in the companion article \cite{Heller:2011ju}.

The equations of hydrodynamics written in the form \eqref{eq.Fw} allow for a simple criterion for thermalization by measuring dimensionless deviation of \eqref{wdef} from obeying \eqref{eq.Fw}. In particular, in \cite{Heller:2011ju} we adopted the following criterium for thermalization
\eq
\label{e.criterion}
\left| \f{\tau \f{d}{d\tau} w}{F_{hydro}^{3^{rd}\, order}(w)} -1
\right| <0.005.
\eqx
Note that the condition \eqref{e.criterion} is based on demanding that the effective temperature obeys equations of hydrodynamics, rather than on isotropy of the pressures. Indeed, as the results of \cite{Heller:2011ju} and earlier studies in \cite{Chesler:2009cy} show, the pressure anisotropy can be quite sizable, nevertheless the evolution of the system being governed by hydrodynamics. In section \ref{sec.profiles} we discuss the sensitivity of thermalization times obtained from \eqref{e.criterion} to the number on the right hand side of \eqref{e.criterion}.

Once gradient terms become negligible, the entropy is no longer changing in time. This can be seen by evaluating the entropy per unit rapidity and transverse area, being a product of the thermodynamic entropy
\eq
s(\tau) = \frac{1}{2} N_{c}^{2} \pi^{2} T_{eff}(\tau)^{3}
\eqx
and the volume element scaling linearly with proper time \eqref{mink.metric.in.bi.coord}. In the following, as in \cite{Heller:2011ju}, in order to measure the final entropy, we will get rid of $N_c^2$ factor as in \eqref{def.effective.temp} and will be using instead dimensionless quantity, being entropy per unit rapidity and transverse area measured in units of the initial effective temperature $T_{eff}^{(i)}$
\eq
s= \f{dS/dy dx_\perp^2}{\frac{1}{2} N_{c}^2 \pi^{2} \big(T_{eff}^{(i)}\big)^2}.
\eqx
Applying this definition in the $\tau \to \infty$ limit of the hydrodynamic
regime gives us an expression for the final (dimensionless) entropy 
expressed in terms of $\Lm$
\eq
\label{entropy.final}
s^{(f)}= \Lm^{2}  \cdot \big(T_{eff}^{(i)}\big)^{-2}.
\eqx
In the latter part of the article we will be using a particular generalization of entropy to non-equilibrium regime $s_{n-eq}$, which however reduces to \eqref{entropy.final} in the regime
of applicability of perfect fluid hydrodynamics.

\subsection{Early time dynamics from holography \label{subsec.earlytimedynamics}}

In the gauge-gravity duality the symmetries of the boundary dynamics are also the symmetries of the dual background, which here is a solution Einstein's equations with negative cosmological constant
\eq
\label{EINSTEINeqn1}
R_{a b} - \frac{1}{2} R g_{a b} -\frac{d(d-1)}{2 L^{2}} = 0
\eqx
with $d$ being the dimension of boundary spacetime taken here to be $1+3$. In the following we set $L$ -- the radius of AdS vacuum solution -- to 1. The most general $(4+1)$-dimensional dual metric sharing the symmetries of the boost-invariant flow takes the form
\eqn
\label{gma}
&& d s_{bulk}^{2} = g_{a b} dx^{a} dx^{b} = \frac{1}{u} \big\{ - A^{2} dt^{2} +t^{2} B^{2} d y^{2} + \nonumber\\ 
&& C^{2} dx_{\perp}^2 + \frac{1}{4 u} D^{2} du^{2} + 2 E dt du \big\},
\eqnx
where the warp-factors are functions of bulk proper time $t$ and radial coordinate $u$ with $u = 0$ denoting the boundary \cite{Buchel:2008ac}. Note that $t$ does not need to coincide with $\tau$ even at the boundary.
The use of $u$ instead of $z = \sqrt{u}$ softens singularities of Einstein's equations at $z=0$, which is convenient when solving them numerically. There is a redundancy in the metric (\ref{gma}) coming from diffeomorphisms in $t$ and $u$, which can be used to fix two out of three warp-factors within $A$, $D$ and $E$. Various choices define various coordinate frames covering different parts of the underlying manifold in various ways. 

The choice $D = 0$ and $E = -\frac{1}{2 \sqrt{u}}$ leads to coordinates of ingoing Eddington-Finkelstein type used in numerical simulations of \cite{Chesler:2008hg,Chesler:2009cy,Chesler:2010bi} (typically the Eddington-Finkelstein coordinates are written using $r = 1/\sqrt{u}$). One reason why in this work we are not using the Eddington-Finkelstein coordinates is that there seems to be a subtlety there in starting at $\tau = 0$ at the boundary (being also $t=0$ in the bulk). This can be seen by looking at the vacuum AdS metric in the Eddington-Finkelstein coordinates
\eq
ds^2 = - \frac{1}{u} dt^2 + \left(1 + \frac{t}{\sqrt{u}} \right)^2 dy^{2} + \frac{1}{u} dx_{\perp}^{2} - \frac{1}{u^{3/2}} dt du,
\eqx
for which ($dy^2$ term) near-boundary limit ($u \rightarrow 0$) does not commute with small time limit $t \rightarrow 0$. The authors of \cite{Chesler:2009cy} have not encountered this problem as their numerical simulations were always starting at $t_{ini} = \tau_{ini} > 0$.

Setting $D = 1$ and $E = 0$ in the metric Ansatz \eqref{gma} leads to the Fefferman-Graham chart, which has been used extensively to understand the asymptotic properties of bulk spacetime relevant for obtaining the stress tensor using the procedure of holographic renormalization \cite{de Haro:2000xn}. In particular, the boundary condition setting the metric in which dual field theory lives to be flat is just
\eq
\label{flatB}
A_{FG} = 1, \quad B_{FG} = 1 \quad \mathrm{and} \quad C_{FG} = 1
\eqx
and in that case one can also identify bulk $t$ and boundary (physical) proper time $\tau$. In the Fefferman-Graham coordinates the near-boundary expansion of warp-factors turns out to occur in integer powers of $u$.  The first subleading term vanishes automatically and the first nontrivial term in the expansion of $A_{FG}$ upon holographic renormalization \cite{de Haro:2000xn} comes out proportional to the energy density of dual $\mathcal{N}=4$ super Yang-Mills plasma $\eps(\tau)$
\eqn
\label{enDEN}
A_{FG} &=& 1 - \frac{\pi^{2}}{N_{c}^{2}} \eps \, u^2 - \frac{\pi^{2}}{4 N_{c}^{2}} ( \frac{1}{\tau} \eps' + \frac{1}{3} \eps'' ) u^3 + \ldots \nonumber\\
B_{FG} &=& 1 - \frac{\pi^{2}}{N_{c}^{2}} (\eps +\tau \eps') u^2 - \frac{\pi^{2}}{3 N_{c}^{2}} ( \eps'' + \frac{1}{4} \tau \eps''' ) u^3 + \ldots \nonumber\\
C_{FG} &=& 1+ \frac{\pi^{2}}{N_{c}^{2}} (\eps +\frac{1}{2} \tau \eps') u^2 + \nonumber \\
&& + \frac{\pi^{2}}{8 N_{c}^{2}} ( \frac{1}{\tau} \eps' + \frac{5}{3} \eps'' + \frac{1}{3} \tau \eps''' ) u^3 + \ldots
\eqnx
$N_{c}$ in the formulas above denotes the number of colors, which although formally infinite, cancels out with $N_{c}^{2}$ contribution in the energy density yielding in the end a finite result. All the remaining terms in the near-boundary expansion of warp-factors turn out to be entirely specified in terms of the energy density and its derivatives. It is worth noting that one way of making sure that dual spacetime written in other coordinate frames has a flat boundary (i.e. imposing such boundary condition), as well as obtaining energy density of dual field theory configuration, is to perform a near-boundary coordinate change to the Fefferman-Graham chart and then directly use (\ref{flatB}) and (\ref{enDEN}) to identify the relevant terms. This will become important later in the article.

The starting point of the analysis of the early time dynamics in \cite{Beuf:2009cx} was the expansion (\ref{enDEN}). Because of the appearance of inverse powers of proper time in terms containing odd derivatives of energy density, demanding finiteness of near-boundary warp-factors in the limit $\tau \rightarrow 0$ gave a highly nontrivial constraint on the early time Taylor series of energy density, so that only even powers of proper time are allowed
\eq
\label{endenexpansion}
\eps(\tau){\big|}_{\tau \approx 0} = \frac{3}{8}N_c^{2} \pi^{2} \left(T_{eff}^{(i)} + \frac{1}{2} T_{eff}''(0) \tau^{2} +\ldots \right)^{4}.
\eqx
One consequence of this is that first proper time derivatives of warp-factors at $\tau = 0$ vanish, so that the two gravitational constraint equations (the $uu$ and $\tau u$ components of Einstein's equations) provide \emph{solvable} relations between $A_{FG} (\tau = 0, u) = A_{0}(u)$, $B_{FG} (\tau = 0, u) = B_{0}(u)$ and $C_{FG} (\tau = 0, u) = C_{0}(u)$ and their radial derivatives (since $\tau$ derivatives vanish at $\tau=0$). One of these relation can be explicitly solved giving
\eq
B_{0} (u) = A_{0} (u),
\eqx
whereas the other leads to a differential equations
\eq
\label{constr2}
\frac{A_{0}''(u)}{A_{0} (u)} + \frac{C_{0}''(u)}{C_{0} (u)} = 0.
\eqx
This equation, via taking suitable combination of $A_{0}$ i $C_{0}$ can be solved exactly (i.e. in terms of a single function specifying initial condition, but with no integration involved), as shown in \cite{Beuf:2009cx}, but for the purposes of this article, it will be sufficient to solve \eqref{constr2} numerically for $A_{0}$ regarding $C_{0}$ as an initial condition. This way of viewing \eqref{constr2} has the advantage that by passing to another coordinate frame in which initial time hypersurface coincides with Fefferman-Graham one, $C_{0}$ will not transform (up to redefinitions of what is meant by $u$). An interesting feature of the equation (\ref{constr2}) is that for all allowed $C_{0}(u)$ there exists a point $u_{0}$ such that $A_{0}(u)$ has a single zero there \cite{Beuf:2009cx}. This point signals the breakdown of the Fefferman-Graham chart and makes it hard, if not impossible, to evolve Einstein's equations in this coordinate frame. A typical, analytic example of initial conditions derived in \cite{Beuf:2009cx} is
\eq
\label{typicaldata}
A_{0} = \cos{\gamma \, u} \,\,\, \mathrm{and} \,\,\, C_{0} = \cosh{\gamma \, u},
\eqx
where $\gamma = \frac{1}{2}\sqrt{3}\pi^{2} \big(T_{eff}^{(i)}\big)^2$ sets the initial energy density and $u$ runs from $0$ to $u^{(FG)}_{0} = \pi / 2 \gamma$ with the Fefferman-Graham coordinate frame breaking down at the latter point. Table \ref{tab:profs} in the appendix summarizes 29 different $C_{0}(u)$ and corresponding values of $u^{(FG)}_{0}$ considered in this article.

Although the near-boundary expansion of $C_{0}$'s stays in an one-to-one correspondence with the early time Taylor series for the energy density \eqref{endenexpansion}
\eq
\label{C0at0}
C_{0}(u) = 1+ \frac{3}{8}\pi^{4} \big(T_{eff}^{(i)}\big)^{4} u^{2} + \frac{1}{2} \pi^{4} \big(T_{eff}^{(i)}\big)^3 T_{eff}''(0) u^{3} + \ldots
\eqx
and given $C_{0}$ in an analytic form (at least close to the boundary) one can work out the first couple of dozen terms in the expansion 
of $\eps(\tau)$, the radius of convergence of the resulting series is insufficient to observe the transition to hydrodynamics (see section \ref{sec.profiles} for an explicit comparison of the effective temperature given by the early time power series with the full numerical result). Not surprisingly, in order to understand thermalization in this model, a full numerical solution of the initial value problem on the gravity side is needed. 

A naive attempt at numerically solving Einstein's equations in the 
Fefferman-Graham frame is bound to fail for two reasons. Firstly, the
generic vanishing of $A_0(u)$ leads to a coordinate singularity.
Secondly, even if this was overcome, some sensible initial conditions
exhibit a curvature singularity in the bulk\footnote{Such initial conditions may be physical if there is an event horizon cloaking the curvature singularity. This will turn out to be the case for the initial conditions considered in the present work.} (at $u\to \infty$).
 
The first step towards solving numerically Einstein's equations is 
thus to find a coordinate frame, which allows to use initial data found in \cite{Beuf:2009cx}, bypasses the breakdown of Fefferman-Graham coordinates
and makes it possible to introduce bulk cutoff for numerical simulation
in order to avoid (a possible) curvature singularity.

\section{New coordinates\label{sec.newcoord}}

The way we deal here with the singularity of Fefferman-Graham frame at $\tau = 0$ is analogous to what happens in the case of Schwarzschild black hole \eq
ds^2 = - (1- \frac{2 M}{r}) dt^2 + (1- \frac{2 M}{r})^{-1} dr^2 + r^2 d\Omega_{S^{2}}^{2}
\eqx
when going from Schwarzschild to Kruskal-Szekeres coordinates. The latter ones are defined by
\eqn
\label{KruskalSzekeresTrafo}
V &=& (\frac{r}{2M} -1)^{1/2} e^{r/4M} \sinh{\frac{t}{4M}} \label{eqV}, \\
U &=& (\frac{r}{2M} -1)^{1/2} e^{r/4M} \cosh{\frac{t}{4M}} \label{eqU}
\eqnx
and lead to manifestly regular metric at $r = 2 M$
\eq
ds^2 = \frac{32 M^3}{r} e^{-r / 2 M} (-dV^{2} + dU^{2}) + r^2 d\Omega_{S^{2}}^{2}
\eqx
with $r$ defined implicitly by the relation $V^{2} - U^2 = (1 - \frac{r}{2 M}) e^{r/2 M}$. What is important for our purposes is that the hypersurface $t = 0$ coincides with the one given by $V = 0$, as follows from \eqref{eqV}. In that way we can take $t = 0$ metric functions at $d\Omega_{S^{2}}^{2}$ (analogues of $dy^2$ and $x_{\perp}^{2}$ warp factors) and use them directly as $d\Omega_{S^{2}}^{2}$ (in our case $dy^2$ and $x_{\perp}^{2}$) warp-factors setting initial data for the evolution in a chart without spurious coordinate singularities (here the analogue of Kruskal-Szekeres coordinates). To make the analogy even sharper, on the $V = 0$ hypersurface one can use $r$ instead of $U$ coordinate at the price that Schwarzschild metric will become $V$-dependent.

Let's try to apply the same logic to the boost-invariant metric in the Fefferman-Graham coordinates
\eqn
\label{FGmetric}
ds^2 = && - \frac{1}{u} A_{FG}(t,u) dt^2 +\frac{t^2}{u} B_{FG}(t,u) dy^2  \nonumber \\
&&+ \frac{1}{u}C_{FG}(t,u) dx_{\perp}^{2} + \frac{1}{4u^{2}} du^{2}
\eqnx
in the neighborhood of $t = 0$ hypersurface. The coordinate singularity arises from $A_{FG}(t = 0, u) = B_{FG}(t = 0, u)$ vanishing at some radial position, whose precise value depends on the choice of initial condition. 
By doing a local coordinate transformation in the neighborhood of $t=0$ (i.e. perturbatively in $t$) one can redefine $t$ in a fashion similar to \eqref{KruskalSzekeresTrafo} and give an arbitrary form to 
the $dt^2$ metric coefficient at this particular instance of time\footnote{An explicit example of such a redefinition is \[t=\f{f(u)}{\sqrt{A_{FG}(0,u)}} \tilde{t} + \oo{\tilde{t}^2}\].}. 

This statement pesists to an arbitrary order in small $t$ expansion and can be understood as adopting a different gauge within the choices allowed by the metric Ansatz \eqref{gma} -- one in which one fixes $A$ and $E$ at the cost of leaving $D$ dynamical. This is advantageous for us, as we can use now ADM formalism-based scheme for numerically integrating Einstein's equations. Moreover, the freedom of choice of $A$ allows to introduce a very natural bulk cutoff as anticipated in the introduction and elaborated on in the next section. Last, but not least, $t=0$ hypersurfaces in both coordinate frames are by definition the same and one can use $u$ as a radial coordinate on both of them. This allows us to start with initial conditions derived earlier in the Fefferman-Graham coordinates and continue the early time power series
solution of \cite{Beuf:2009cx} beyond its radius of convergence in order 
to explore the transition to hydrodynamics, which was one motivation 
for the present work.

\section{Review of ADM formulation of General relativity\label{sec.admreview}}

A particular formulation of Einstein's equations which is very convenient
for studying evolution from generic initial data is the ADM formulation~\cite{MWT-ADM,Shibata:2009ax}.
In this work we did not adopt any of its more refined versions like
BSNN~\cite{Shibata:2009ax}, as in 1+1 (and even 2+1) ordinary ADM should suffice. 

The key idea behind the ADM formalism, making it at the same time a natural point of departure in implementing Einstein's equations numerically, is to assume that there exists a global 
foliation of spacetime into spacelike hypersurfaces of constant time 
and recasting Einstein's equations in terms of equations intrinsic to a constant time slice (constraint equations) and extrinsic ones (encoding the time-evolution) \cite{poissonbook}. As it will turn out, due to the choice of coordinates we will be using foliations of \emph{patches} of the bulk spacetime, rather than global foliations. 

Denoting by $\lambda$ a scalar function slicing the bulk onto constant time hypersurfaces, the induced metric describing the intrinsic geometry of a leaf takes the form 
\eqn
\label{projector}
\gamma_{a b}=g_{a b}+n_a n_b,
\eqnx
where $n^{a}$ is a future directed unit normal vector obtained from the gradient of $\lambda$. The induced metric \eqref{projector} acts also as the projector operator onto foliation leaves. The spacetime embedding of the constant time hypersurface is described by the extrinsic curvature 
\eqn
K_{a b}=-\frac{1}{2}\LL_n\gamma_{a b},
\eqnx
where $\LL_n$ denotes the Lie derivative along the normal direction $n^{a}$. A coordinate basis temporal vector, can be constructed from the unit normal $n^a$ as
\eqn
\frac{\partial x^{a}}{\partial \lambda}=t^a=\alphat n^a+\beta^a
\eqnx
with $\alphat$ and $\beta^{a}$ being respectively the lapse and the shift vector ($n^{a}\beta_{a} = 0$). The role of the lapse is to measure the rate of time flow between subsequent slices of the foliation, whereas the shift vector describes how the hypersurfaces slide onto each other in transverse directions. With the use of the projector \eqref{projector}, Einstein's equations
\eqn
\label{EINSTEINeqn2}
R_{a b} - \frac{1}{2} R \, g_{a b}=8\pi G_{N} \, T_{a b}
\eqnx
can be decomposed into constraints and evolution equations. The bulk energy-momentum tensor (not to be confused with the expectation value of the boundary stress tensor operator!) is decomposed into density, current and a transverse tensor taking respectively the form
\eqn
\rho=T_{a b}n^a n^b,\ j_c=-T_{a b}n^a\gamma^b_{c},\ S_{c d}=T_{a b}\gamma^a_{\ c}\gamma^b_{\ d}.
\eqnx
In the ADM formulation, equations governing internal spacelike geometry of the hypersurface are recast in the form of constraint equations reflecting the time and space decomposition of spacetime. The Hamiltonian constraint, following from Gauss equation, reads
\eqn
R+K^2-K_{a b}K^{a b}=16\pi G_{N} \rho,
\eqnx
whereas the momentum constraint derived from Codazzi equation is
\eqn
D_{b}K^b_{\ a}-D_{a}K=8\pi G_{N} j_a.
\eqnx
$D_a$ here is the covariant derivative compatible with the spatial metric $\gamma_{a b}$ and $\cal R$ is the Ricci scalar associated with it. Evolution equations for the induced metric come from projecting its Lie derivative onto constant time slice
\eqn
\LL_t \gamma_{a b}=-2\alphat K_{a b}+D_a \beta_b+D_b \beta_a.
\eqnx
The evolution equations for the external curvature can be obtained in a similar fashion from the Ricci equation
\eqn
\LL_t K_{a b} &=& -D_a D_b\alphat+\alphat({\cal R}_{a b}-2K_{a c}K^c_{\ b}+K_{a b}K) \nonumber\\
&+& \beta^c D_c K_{a b}+K_{c b}D_a \beta^c+K_{c a}D_b\beta^c \nonumber\\
&-& 8\pi G_N\alphat[S_{a b}+\frac{\rho-S}{d-1}\gamma_{a b}].
\eqnx
Here ${\cal R}_{a b}$ is the Ricci tensor of $\gamma_{a b}$, $S=S^a_{\ a}$ and $d$ denotes dimensionality of boundary (taken everywhere in the paper, apart from this section where it is kept general, to be $1+3$). One can notice that the only place where the bulk stress tensor contributes is the last term and this is where vacuum AdS cosmological constant will manifest itself. Comparing \eqref{EINSTEINeqn1} with \eqref{EINSTEINeqn2} one can see that in the absence of matter fields the bulk energy-momentum tensor $T_{a b}$ is related to the radius of vacuum AdS solution $L$, $(d+1)$-dimensional Newton's constant $G_{N}$ and bulk metric $g_{a b}$ through
\eqn
T_{a b} = \frac{d(d-1)}{16 \pi G_N L^{2}} \, g_{a b}.
\eqnx
By introducing transverse coordinates of the foliation leaf $y^i$ one can recast the metric into the standard ADM form
\eqn
\label{metricADM}
ds^2=-\alphat^2d\lambda^2+\gamma_{ij}(dy^i+\beta^i d\lambda)(dy^j+\beta^j d\lambda).
\eqnx
By doing so one can confine $(d+1)$-dimensional indices to $d$ spatial slices as in the projected quantities normal components are zero (we shall denote those by $i$ and $j$ Latin letters, as opposed to $a$, $b$ and $c$ symbols running through $d+1$ indices). Moreover, now Lie derivative along $n^{a}$ is simply a time derivative $\partial_\lambda$ (since $\lambda$ parametrizes curves normal to slices).

\section{ADM formulation for boost-invariant plasma\label{sec.admforbif}}

Before we apply the ADM equations in the context of the time evolution of
the boost-invariant geometry in an asymptotically AdS geometry, we have to 
specify certain additional ingredients. 

Firstly, due to the fact that 
the metric blows up at the AdS boundary, one has to redefine the
ADM variables i.e. the spatial metric coefficients and the extrinsic curvature
by taking out predefined factors which will ensure that the asymptotic behaviour
is fulfilled while keeping all the new redefined variables finite throughout the integration
domain. 

Secondly, we have to specify the initial conditions which satisfy constraint
equations. In the special case of a boost-invariant geometry with initial conditions
set at $t=0$, this requires some care as the initial hypersurface is not
strictly spacelike but has signature $(0,+,+,+)$. This feature simplifies
the determination of possible consistent initial conditions, but also requires
special treatment of the first step of the temporal evolution. 
Of course, for all $t>0$, the constant $t$ hypersurfaces are spacelike and
conventional ADM formulation applies.

Thirdly, we have to impose boundary conditions for all dynamical variables
at the AdS boundary. 
These boundary conditions are conceptually clear, as they correspond to 
enforcing a Minkowski metric on the boundary (in order to ensure that 
dual $\nn=4$ SYM theory is defined on an ordinary Minkowski space). However,
it turns out that for a generic choice of lapse functions, the boundary metric
may be related to Minkowski by a boundary diffeomorphism. Taking this into
account leads to more complicated boundary conditions than conventional Dirichlet
boundary conditions. 

Fourthly, we have to impose boundary conditions at the outer edge of 
the integration domain in the bulk. These boundary conditions are much more 
subtle and are not fixed by the very principles of AdS/CFT (as are the previous
boundary conditions) but rather depend on the specific features of the dynamical
problem at hand. The main requirements are that these boundary conditions should
not interfere with the physics of interest and also should lead to stable numerical
evolution.

Finally, we have to specify the final ingredients of the ADM formalism -- the lapse
and shift functions which encode how the hypersurfaces of fixed (simulation-)time
fit together to form the 5D geometry. For our purposes we set the shift vector 
to zero, however the lapse will remain nontrivial. A part of the lapse function
will be used in defining the outer edge boundary conditions but the remaining part
will have to be specified.

In the following we will discuss in turn all the above mentioned points.

\subsection{Rescaled ADM variables and equations of motion}

In order to ensure that the functions entering the ADM equations are 
finite everywhere, even at the AdS boundary, we have factored out appropriate factors
of $u$ (the AdS bulk coordinate). Our $(4+1)$-dimensional metric takes the form
\eqn
\label{e.5d}
ds^2 &=& \frac{-a^2(u) \alpha^2(t,u) dt^2}{u}+ \frac{t^2 a^2(u) b^2(t,u) dy^2}{u} \nonumber\\
 &+&\frac{c^2(t,u) dx^2_\perp}{u}+\frac{d^2(t,u) du^2}{4u^2}
\eqnx
with empty AdS being represented by all the coefficient functions equal to unity.
Note that in general the time coordinate $t$ will not be equal to the gauge theory
proper time $\tau$ at the boundary. We will derive an explicit relation between
the two coordinates in subsection \ref{subsec.bdrycondAdSbdry}.
With the above definition, the ADM spatial metric $\gm_{ij}$ is given by
\eq
\gm_{ij}=diag\left[
\frac{t^2 a^2(u) b^2(t,u)}{u}, \frac{c^2(t,u)}{u},
\frac{c^2(t,u)}{u}, \frac{d^2(t,u)}{4u^2} \right].
\eqx
The nontrivial components of the extrinsic curvature are also rescaled
\eq
K_{ij}=diag\left[\frac{t a(u) L(t,u)}{\sqrt{u}},\frac{M(t,u)}{\sqrt{u}},
\frac{M(t,u)}{\sqrt{u}},\frac{P(t,u)}{4u\sqrt{u}}\right].
\eqx
Finally, the lapse function is parametrized by
\eq
\alt(t,u)=\f{a(u) \al(t,u)}{\sqrt{u}}.
\eqx
The reason of factoring out the time independent function $a(u)$ will be clear when
we discuss the outer boundary conditions below. For simplicity we will always set
$a(0)=1$.

In the presence of a cosmological constant $\Lm=-6$ and vanishing shift the vacuum
ADM equations become
\eqn
&\partial_t& \gamma_{ij}(t,u)=\frac{-2a(u)\alpha(t,u)}{\sqrt{u}}K_{ij}(t,u), \label{eq:ADM1} \\
&\partial_t& K_{ij}=-\nabla_i\nabla_j \frac{a(u)\alpha(t,u)}{\sqrt{u}}+4\frac{a(u)\alpha(t,u)}{\sqrt{u}}\gamma_{ij}(t,u),\nonumber\\ 
&+&\frac{-2a(u)\alpha(t,u)}{\sqrt{u}}(R_{ij}-2K^{ij}K_{ij}+KK_{ij}). \nonumber
\eqnx
For the metric coefficients we obtain:
\eqn
\frac{\partial b(t,u)}{\partial t}&=&\frac{-b(t,u)^2+\alpha(t,u)L(t,u)}{t b(t,u)}, \label{eq:ADM2} \\
\frac{\partial c(t,u)}{\partial t}&=&\frac{a(u)\alpha(t,u)M(t,u)}{c(t,u)}, \nonumber\\
\frac{\partial d(t,u)}{\partial t}&=&\frac{a(u)\alpha(t,u)P(t,u)}{d(t,u)}. \nonumber
\eqnx
The evolution equations for the extrinsic curvature functions $\partial_t L,\ \partial_t M,\ \partial_t P$ are quite lengthy and can be found in Appendix \ref{app:b}. They were generated by {\it Mathematica} and directly transfered to the computer code.

The hamiltonian and momentum constraints take the form
\eqn
\label{e.constraints}
&C_0& = R-K_{ij}K^{ij}+K^2+12, \nonumber\\
&C_i& =  \nabla_{i}(K^{ij}-K\gamma^{ij}),
\eqnx
with only the $u$ component of the momentum constraint $C_4$ being \emph{a-priori}
nontrivial.

\subsection{Initial conditions at $t=0$ and the first step of the numerical evolution}

Typically in the ADM formalism, the initial conditions are obtained by solving
the constraint equations (\ref{e.constraints}). In the case of the boost-invariant
geometry, however, the initial hypersurface $t=0$ is not spacelike as it contains
the light cone in the longitudinal plane. Its signature is $(0,+,+,+)$. This 
requires an \emph{ab-initio} analysis of initial conditions, which become in fact
unconstrained, and a special treatment of the evolution equations (i.e. the right hand side
of $\partial_t b$, $\partial_t c$ etc.) at $t=0$.

To this end we will expand the lapse and all coefficient functions up to linear
order in $t$:
\eqn
\label{e.smallt}
b=b_0(u)+b_1(u)t+\ldots &\quad& c=c_0(u)+c_1(u)t+\ldots \nonumber\\
d=d_0(u)+d_1(u)t+\ldots &\quad& L=L_0(u)+L_1(u)t+\ldots \nonumber\\
M=M_0(u)+M_1(u)t+\ldots &\quad& P=P_0(u)+P_1(u)t+\ldots \nonumber\\
\al=\al_0(u)+\al_1(u)t+\ldots &&
\eqnx
After inserting these expansions into both the ADM dynamical equations (\ref{eq:ADM1})
and the constraints (\ref{e.constraints}), we obtain the initial conditions 
($b_0,c_0$ etc.) as well as evolution equations at $t=0$. We found by an explicit calculation that
both $c_0(u)$ and $d_0(u)$ are unconstrained and free, whereas $b_0(u)$ turns out to be
proportional to $\al_0(u)$. Without loss of generality we will set the constant
of proportionality to unity. All this taken into account gives
\eq
b_0(u)=\al_0(u) \quad L_0(u)=\al_0(u) \quad M_0(u)=P_0(u)=0.
\eqx
Note that we are free to perform a spatial diffeomorphism (redefine $u$). In this way
we may freely set $d_0(u)=1$, leaving the initial condition to be completely specified
by a single function $c_0(u)$. In the following we will restrict
our lapse functions to satisfy $\al_0(u)=1$. With these choices made, the final
initial conditions for the ADM evolution are
\eqn
\label{e.init}
b_0(u)=d_0(u)&&=L_0(u)=1 \quad M_0(u)=P_0(u)=0 \nonumber\\ 
&&c_0(u)\equiv c_0^{profile}(u)
\eqnx
parametrized by the single function $c_0^{profile}(u)$. Comparing the resulting
initial geometry with the Fefferman-Graham initial condition we see that we can identify
$c_0^{profile}(u)$ with the Fefferman-Graham initial condition $C_{FG}(\tau=0,u)$
discussed in subsection \ref{subsec.earlytimedynamics}. Consequently, we have at our disposal a power series
solution for $\eps(\tau)$, which may be used to check the results of the numerical
evolution for some initial range of $\tau$.

In some cases, when running the simulations, we noticed that quite narrow
structures emerge at late times close to the outer edge of the simulation domain
in the bulk. This causes the numerical evolution to eventually break down for
a given size of the spatial grid. In these cases we found it useful to redefine
the initial coordinates by
\eq
u \to \f{u}{1-C u}
\eqx
with appropriately choosen constant $C$. This redefinition stretches the grid
at the outer edge allowing in some cases for longer evolution. The initial
conditions now take the form
\eqn
\label{e.init2}
d_0(u)&&=\f{1}{1+C u} \quad c_0(u)=\sqrt{1+C u} \cdot
c_0^{profile} \left( \f{u}{1+C u} \right), \nonumber\\
&&b_0(u)=L_0(u)=1 \quad M_0(u)=P_0(u)=0  .
\eqnx
Of course all the physics extracted from running the simulation from the initial
conditions (\ref{e.init}) and (\ref{e.init2}) with the same function $c_0^{profile}(u)$
is completely equivalent.

The terms linear in $t$ in (\ref{e.smallt}) give the right hand sides 
of the evolution equations at $t=0$, which we use for the first time-step
of the numerical integration.
We obtain in this way
\eqn
&&\partial_t b(0,u) = 0,\quad\quad \partial_t c(0,u)=0, \\
&&\partial_t d(0,u) = 0,\quad\quad \partial_t L(0,0)=0, \nonumber \\
&&\partial_t M(0,u) = \frac{-2\alpha(2u(c\partial_uc(a\alpha)+\partial_u(a\alpha c\partial_uc))}{d^2} \nonumber \\
&&+\frac{a\alpha\partial_ud(c^2+\partial_uc^2)}{d^3}-\frac{2a\alpha c^2}{u}+\frac{2a\alpha c^2}{ud^2}, \nonumber \\
&&\partial_t P(0,u) = 4u\partial^2_u(a\alpha)+\frac{4\partial_ud(a\alpha+\partial_u(a\alpha))}{d} \nonumber \\
&&-\frac{2a\alpha(1-d^2)}{u}+\frac{4ua\alpha\partial^2_uc}{c}-\frac{4ua\alpha\partial_uc\partial_ud}{cd}. \nonumber
\eqnx

As a final note let us clarify why in the present ADM formalism we have completely
unconstrained initial data, while in the Fefferman-Graham case we had the differential
constraint (\ref{constr2}). The Fefferman-Graham coordinates are a special case of
the general metric ansatz (\ref{e.5d}) albeit with the constraint that $d(t,u) \equiv 1$
(imposing this condition, however, transforms the lapse into a dynamical variable).
Requiring that the condition $d(t,u)=1$ is preserved under evolution requires that
$P(t,u)\equiv 0$ and in particular $\partial_t P(0,u)=0$. It is this last equation 
which reduces exactly to the the Fefferman-Graham initial data 
constraint (\ref{constr2}).

\subsection{Boundary conditions at the AdS boundary\\ at $u=0$\label{subsec.bdrycondAdSbdry}}

The boundary conditions at $u=0$ are choosen as to ensure that the gauge
theory metric is just the $(3+1)$-dimensional flat Minkowski metric. The easiest way
to derive these conditions is to start from the leading asymptotics in 
Fefferman-Graham coordinates (i.e. basically the empty $AdS_5$ geometry)
\eq
\label{e.fgempty}
ds^2=\f{-d\tau_{FG}^2 +\tau_{FG}^2 dy^2+dx^2_\perp+dz^2}{z^2}
\eqx 
and consider the most general change of coordinates (in an expansion around $u=0$)
to our ADM metric ansatz (\ref{e.5d}). Hence we write
\eqn
\label{e.transform}
\tau_{FG} &=& f_0(t)+f_1(t) u+ f_2(t) u^2+\ldots \nonumber\\
z &=& g_0(t)u^{\f{1}{2}}+g_1(t) u^{\f{3}{2}}+g_2(t) u^{\f{5}{2}}+\ldots
\eqnx
The physical gauge theory proper time is just $\tau=f_0(t)$.
Inserting this into (\ref{e.fgempty}) we see at once from the $du^2$ component
that the boundary condition for $d$ is just 
\eq
d(t,0)=1.
\eqx
Consequently $P(t,0)=0$.
However the boundary values of $b$ and $c$ can be $t$-dependent. 
We get that $g_0(t)=1/c(t,0)$ (from $dx^2_\perp$) and combining this with
\eq
\f{f_0}{g_0}=a(0) b(t,0) t
\eqx
(from $dy^2$) we obtain a very important relation between the coordinate time $t$ in our ADM
simulation and the physical proper time (recall that we have $a(0)=1$)
\eq
\tau \equiv f_0(t)=\f{b(t,0) t}{c(t,0)}.
\eqx
The condition that ensures having the Minkowski metric will be the compatibility
of the above equation for $f_0(t)$ with last remaining condition coming from the
$dt^2$ component of the metric
\eq
\label{e.f0dot}
\dot{f_0}=\f{\al(t,0)}{c(t,0)}.
\eqx
This condition leads to an expression for the boundary value of $L$
\eq
L(t,0)=b(t,0)+t \f{b^2(t,0)}{c^2(t,0)} M(t,0).
\eqx
For actual numerical implementation it is convenient to impose the above boundary
condition in a differential form
\eq
\partial_t L=\partial_t \left( b+t \f{b^2}{c^2} M \right).
\eqx
At this stage we are missing only the evolution equation for $M(t,0)$.
There are various ways of implementing this equation. Perhaps the simplest
way is to expand the evolution equation for $\partial_t M$ around $u=0$ and
take into account the conditions derived above i.e. $d(t,0)=1$, $P(t,0)=0$ and
the expression for $L(t,0)$.

An alternative way is to compute the first subleading terms in the coordinate 
transformation (\ref{e.transform}). In any case this computation has to be done in
order to derive the formula for the energy density $\eps(\tau)$.
Requiring that there is no $dudt$ term, one can show that 
\eq
f_1(t)=\f{g_0(t) \dot{g_0}(t)}{2 \dot{f_0}(t)}.
\eqx
$g_1(t)$ can be computed by looking at the first subleading term in the $dt^2$
component of the metric and can be expressed in terms of $\al$, $\partial_u \al$,
$\partial_u a$, as well as $f_0(t)$, $g_0(t)$ and their derivatives. Finally, we
may use these expressions to compute $\partial_u d$ (at $u=0$) from the first subleading 
term in the $du^2$ component.
After taking into account all the above definitions and relations, the result may
be brought to the form (all expressions are evaluated at $u=0$)
\eq
\partial_u d=-2\du a -\f{2 \du \al}{\al} +\f{3M^2}{2c^4}- \f{\dt M}{\al c^2}.
\eqx 
This equation allows us to solve for $\dt M$ which is the last remaining equation
at the AdS boundary.

To summarize, the boundary conditions at $u=0$ are formulated as the following evolution equations for the boundary values of the ADM variables
\eqn
\label{e.bndryeom}
&&\partial_t b(t,0) = \frac{-b^2+\alpha L}{t b}, \nonumber \\
&&\partial_t c(t,0) = \frac{\alpha M}{c}, \nonumber \\
&&\partial_t d(t,0) = 0, \nonumber \\
&&\partial_t L(t,0) = \partial_t(b+t\frac{b^2}{c^2}M), \nonumber \\
&&\partial_t M(t,0) = -2c^2(\partial_u (a\alpha)+\alpha\partial_ud)+\frac{3}{2}\frac{\alpha M^2}{2c^2}, \nonumber \\
&&\partial_t P(t,0) = 0.
\eqnx

As a final point let us comment on the regularity of the ADM evolution equations in the bulk as we approach $u \to 0$. The evolution equations
for the metric coefficients are explicitly regular. The right hand side of
the evolution equations for the extrinsic curvature coefficients have,
however, the following structure
\eq
\partial_t \{L,M,P\} =  \f{1}{u} \left( d^2-1 \right) \cdot finite +regular.
\eqx
Since our boundary condition for $d$ is $d(t,0)=1$, the above term does
not lead to any numerical problems. This is partly because we use
spectral methods which preserve very well the smoothness of functions close 
to the boundary. 

\subsection{Boundary conditions at the outer edge of the simulation domain \label{subsec.outerbc}}

As explained in section II, there is a subset of initial conditions for which
the curvature goes to infinity in the center of $AdS$. These initial conditions
still may be perfectly physical if the singularity would be surrounded by 
an event horizon. However \emph{a-priori} we do not know where the event horizon
is located. We can locate it only {\it after} running a simulation.
Moreover due to the null character of the initial hypersurface at $\tau=0$,
the conditions for an apparent (dynamical) horizon cannot be met there.

In order to perform the simulation we have to cut off our numerical grid at some
finite value of $u=u_0$ and impose boundary conditions which would not contaminate
the true physical evolution. The more or less standard choices in numerical
relativity cannot be applied here. Putting any boundary conditions inside the event
horizon cannot be used in our case as the location of the event horizon is
unknown when we start the simulation. Moreover, the spectral discretization
that we use makes the setup very sensitive to the boundary conditions
since Chebyshev differentiation is very much nonlocal in contrast to finite difference
discretization. The second standard choice, namely imposing outgoing radiative
conditions at the outer edge of the simulation domain is also not an option, as
the geometry at the outer edge may be highly curved.

\begin{figure}
\includegraphics[width=5.5cm]{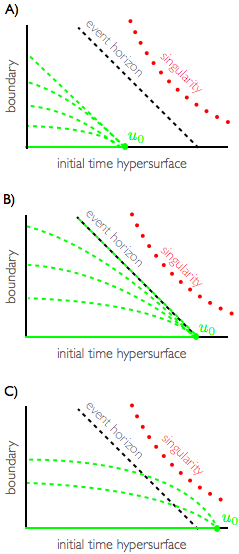}
\caption{Schematic view of constant time foliations depending on the locus at which the lapse vanishes ($u_0$). Situation A) shows a simulation filling in a triangle covering only a finite interval of boundary time. By pushing the position of bulk cutoff inwards (until one passes the event horizon), one can recover the dynamics on larger and larger regions of boundary. In the situation B) the lapse vanishes at the position of the event horizon on the initial time hypersurface, which ensures maximally long simulation time (theoretically infinite) and is optimal for studying the transition to hydrodynamics in dual gauge theory. In the case C) the simulation penetrates into the black brane/black hole interior revealing the apparent horizon at the cost of breaking down when constant time slices start approaching
the curvature singularity. This type of behavior can be seen explicitly in Fig.~\ref{fig:ingoingnullAH} showing the results of a sample simulation.}
\label{fig:u0tuning}
\end{figure} 

In order to bypass the above difficulties, we have decided to use the freedom
of the choice of spatial foliations in the ADM formalism by requiring that
all hypersurfaces (i.e. for any fixed $t$) pass through the \emph{same} single
spacetime point $u=u_0$ on the initial hypersurface $t=0$. This can be done 
technically by freezing the evolution at $u=u_0$ which amounts to forcing the
lapse to vanish there. We achieve this by setting the $t$-independent part of
the lapse in (\ref{e.5d}) to be
\eq
a(u)=\cos\left( \f{\pi}{2} \f{u}{u_0} \right).
\eqx
All ADM evolution equations are regular at $u=u_0$ and indeed do not lead to any
instabilities. Moreover, the region of spacetime outside our numerical grid
is causally disconnected from the simulation domain. Namely, the asymptotic 
boundary of our simulation domain will be a null geodesic running from the
spacetime point $u=u_0$ and $t=0$ in the direction of the boundary. 

However we must note that the optimal choice of $u_0$ is crucial c.f. Fig.~\ref{fig:u0tuning}.
If $u_0$ lies between the event horizon and the boundary, the null geodesic from
$u_0$ will reach the boundary at some time $t=t_*$. Then clearly the ADM
simulation will never be able to go past $t=t_*$ and the simulation will
break down there. The optimal choice would be to set $u_0$ to be exactly
at the position of the event horizon. Then, the simulation could in principle
be run indefinitely and the simulation domain would fill the whole exterior
of the event horizon. This choice is the one we are aiming at in order to extend
the simulation to late times and to observe the transition to hydrodynamics.
Also in this case, the event horizon ensures that the simulation would never
encounter the curvature singularity.
If, on the other hand, $u=u_0$ would lie beyond the event horizon, the simulation
would break down at some finite time due to the curvature singularity. This choice
of $u_0$ is in fact also useful, as it allows us to locate the apparent (dynamical)
horizon and e.g. measure the entropy of the plasma system. This choice
is also crucial in determining the optimal value of $u_0$ as explained in subsection \ref{subsec.AH}
below.

As far as we know, the outer boundary condition obtained by freezing the lapse
has not been used in the literature so far.

\subsection{The choice of lapse functions}

Even after fixing $a(u)$ we still have to determine the final ingredient in the
ADM formulation -- the remaining part of the lapse function $\al(t,u)$. If we were
to adopt the simplest choice $\al(t,u)=1$, then we would encounter coordinate 
singularities (e.g. for the profile $c_0=\cosh \gamma u$, the $d(t,u)$ warp-factor develops a zero).
This is in fact a very well known behaviour of the ADM formalism in asymptotically
flat spacetime \cite{Alcubierre:2002iq,Baumgarte:2002jm,Gourgoulhon:2007ue}, which can be remedied by standard choices of
dynamical lapse functions, like $\sqrt{\det \gm_{ij}}$ or $1+\log\det \gm_{ij}$.
However these choices do not seem to work well in the case at hand. We have chosen
a couple of lapse functions by trial and error. The rationale was e.g. to make 
$\al$ proportional to $d$ to avoid the vanishing of $d$ and make it inversely
proportional to $b$ in order to avoid a too quick rise of $b$.

We always normalized our lapse functions to be equal to 1 at $t=0$. Therefore
we set
\eq
\al(t,u) = \f{lapse(t,u)}{lapse(0,u)}.
\eqx
The choices which seem to work best, i.e. gave long enough bulk evolution to see thermalization in the boundary theory, depending on the initial profile were
\eqn
lapse_1=\f{dc^2}{b}, \quad && lapse_2=\f{b d}{1+\f{u}{u_0} b^2},
\quad
lapse_3=\f{d}{b} \nonumber \\
\mathrm{and} \quad && lapse_4 = d.
\eqnx

Another issue which influenced the choice of lapse functions is that for some
of them one could ensure that the simulation time $t$ coincided with the proper time on the boundary. This happens when $b=c$ at $u=0$. In 
order to preserve this equality under evolution, it is enough to require that\footnote{
This follows directly from equation (\ref{e.f0dot}).} $\al(t,0)=c(t,0)$ (or $b(t,0)$). So from the above lapses, $lapse_1$ 
and $lapse_2$ ensure the equality $t=\tau$ at the boundary.

\section{Observables\label{sec.observables}}

With all the above ingredients in place one can run the numerical evolution,
which is described in more detail in section~VII.
Once the geometry is known, we will be interested in extracting the energy-momentum
tensor from the near-boundary behaviour of the metric.  
Another quantity of interest will be the determination of the 
appearance and the precise location of an apparent horizon and the extraction
of a nonequlibrium entropy of the plasma system.

\subsection{Energy density and transverse and longitudinal pressure\label{subset.formulasfordualstresstensor}}

Obtaining the components of the energy-momentum tensor from the results of the
numerical simulation in the ADM variables turns out to be surprisingly subtle.
The main complication comes from the fact that the radial $du^2$ component of
the metric is a nontrivial dynamical time-dependent field and in addition 
our ADM time coordinate may differ from the physical Minkowski proper time.

The way to perform the computation is to perform the change of variables 
(\ref{e.transform}) from the Fefferman-Graham form
\eqn
ds^2&=&\f{-(1-\f{2\pi^{2}}{N_{c}^{2}}\eps(\tau) u^2) d\tau_{FG}^2 +\tau_{FG}^2 (1+\f{2\pi^{2}}{N_{c}^{2}} p_L(\tau) u^2)dy^2}{u}
+\nonumber\\
&&+\f{(1+\f{2\pi^{2}}{N_{c}^{2}} p_T(\tau) u^2)dx^2_\perp+\f{1}{4u}du^2}{u}
\eqnx  
and compare the result with the near boundary expansion of the ADM metric
(\ref{e.5d}). This requires the computation of the terms $f_2(t)$ and $g_2(t)$
in (\ref{e.transform}) which is straightforward but requires the use of {\it Mathematica}.
We also repeatedly use the boundary equations of motion (\ref{e.bndryeom}).
Then we may extract $\eps(\tau)$, $p_L(\tau)$ and $p_T(\tau)$ from expressions
for $\du^2 d(t,0)$, $\du^2 b(t,0)$ and $\du^2 c(t,0)$ respectively. Note that we
extract directly $p_L$ and $p_T$, even though they could be obtained from $\eps(\tau)$
using (\ref{plANDpt}), for two reasons -- we avoid taking a temporal derivative of
the numerical data for $\eps(\tau)$ and as a byproduct can check whether the numerical
evolution preserves the conservation of the energy-momentum tensor.

After carrying out the calculations mentioned above we arrive at the following
formula for $\eps(\tau)$
\eqn
\eps(\tau)= && \f{N_{c}^{2}}{2\pi^{2}} c^4 \Bigg(-\du^2 a - \f{2 \du a \du \al }{\al}-
\f{\du^2 \al }{\al}+ \du a  \du d+
\nonumber\\
&&+\f{\du \al  \du d}{\al} + \f{3}{4}  \du d^2-
\f{1}{4}  \du^2 d +\f{ \du \dt P}{4\al} \Bigg)
\eqnx
and similar formulas for $p_L(\tau)$ and $p_T(\tau)$. This formula still has a
drawback, as it includes dependence on $\du\dt P(t,0)$. However we can evaluate
this expression by taking the $u$ derivative of the equation for $\dt P(t,u)$
and expanding this near the boundary. In this way we elimated all time derivatives
from the right hand side of $\eps(\tau)$. The final expression for $\eps(\tau)$ is
\eqn
\eps(\tau)= && \f{N_{c}^{2}}{2\pi^{2}} c^4 \Bigg( \du^2 a+\f{2\du a \du b}{b}-\du a \du d- \f{\du b \du d}{b}
- \nonumber\\
&&-\f{2\du c \du d}{c}-\f{9}{4} \du d^2-\f{\du P}{4 t b}
-\f{3M \du P}{4 c^2}+\nonumber\\
&&+\f{\du^2 b}{b}+2 \f{\du^2 c}{c}+\f{3}{4} \du^2 d  \Bigg).
\eqnx
The expression for $p_T$ is given by
\eqn
p_T(\tau) &=& \f{N_{c}^{2}}{2\pi^{2}} \Bigg(c^3 \du^2 c-\f{1}{4}c^4 \du d^2+ \f{1}{4} c^4 \du^2 d 
+\f{1}{4} \du d M^2-\nonumber\\
&& -\f{1}{4} c^2 M \du P \Bigg).
\eqnx
The longitudinal pressure $p_L(\tau)$ can be found from tracelessness 
$\eps(\tau)=p_L(\tau)+2p_T(\tau)$.

The above expressions are quite complex and in order to test them for 
possible errors we have compared $\eps(\tau)$ extracted using the above 
expressions from the numerical simulation with the power series solution for
$\eps(\tau)$ and found complete agreement (see Fig. \ref{fig:series} later in the text).
A further test was to extract $\eps(\tau)$ from two simulations which used
different lapse functions. Again we found agreement.

\subsection{Apparent horizon entropy density}
\label{subsec.AH}

Apart from the energy density and longitudinal and transverse pressures, 
a very important physical property of the evolving plasma system is its 
entropy density per transverse area and unit (spacetime) rapidity. 
According to the AdS/CFT correspondence, gauge theory entropy can be 
reconstructed from the Bekenstein-Hawking entropy of a horizon. This is well 
established and unambiguous in the static case, however in the far-from-equilibrium time-dependent setup the situation is much more subtle.
In fact, it is not even clear whether an exact \emph{local} notion 
of entropy density makes sense on the QFT side. However, 
phenomenological notions
of local entropy density are widely used in dissipative hydrodynamics.

In the present work, we adopt a natural geometrical prescription for local 
entropy density which reduces to the one used in dissipative hydrodynamics
in its regime of validity.

On the AdS side, the definition of entropy density involves two distinct
steps. In the first step, one calculates the area element of a 
(certain kind of a)
horizon. The $0^{th}$ order requirement that one has to impose is that 
the horizon area only increases i.e. the horizon satisfies an area theorem.
Moreover, causality has to be preserved i.e. the horizon area cannot increase
in anticipation of some event. This condition rules out the use of 
event horizons \cite{Chesler:2008hg,Figueras:2009iu}. Currently, the most 
natural choice is the use of apparent horizons described in detail below.

In the second step, one has to link the area element of the horizon to
a specific point on the boundary in order to associate the local
entropy density (area) to a definite spacetime point in gauge theory.
We follow the proposal of \cite{Bhattacharyya:2008xc} of shooting
a null geodesic from the point at the boundary and taking the area
element from the point of intersection of the null geodesic with
the apparent horizon.

Below we discuss these two steps in more detail.

Apparent horizons are quasilocal\footnote{Their definition requires the existence of fully trapped surfaces in the neighborhood of an apparent horizon, hence the term quasilocal.} notions of black hole boundaries, always confined to the causal interior of a black hole 
\cite{Ashtekar:2004cn,Booth:2005qc}. From the point of view of this paper, they are of interest for two reasons. In the first place, their existence is
useful, because an intrinsic (``local'') notion of crossing the boundary of black hole can be utilized to adjust the radial cutoff for integrating Einstein's equations in numerical simulations. 
Secondly, similarly as for event horizons, their area only increases
thus satisfying an area theorem. These properties make them a good
ingredient for a causal generalization of entropy to non-equilibrium situations in holographic systems. All this is in stark contrast with the event horizon with its teleological nature and acausal evolution 
\cite{Booth:2005qc}.

The notion of an apparent horizon, and so entropy defined on it, is tied to a particular foliation of spacetime, which leads to ambiguities. Indeed, in order to find an apparent horizon, one first defines the foliation of spacetime into constant time slices. 
Then on each slice one locates, if it exists, such a codimension-2 hypersurface that wavefronts of light rays emitted (into the future) in an outward direction (i.e. here towards the boundary) stay constant in area, whereas wavefronts emitted in the inward direction shrink. The union of all these codimension-2 hypersurfaces, being itself a codimention-1 hypersurface, is what we call here, with a slight abuse of terminology, an apparent horizon (in fact this is the definition of a future outer trapping horizon). Closer considerations reveal that the area of constitutive slices of apparent horizon is never decreasing, leading to area theorem \cite{Booth:2005qc}.

In highly symmetric spacetimes, such as gravity duals of boost-invariant 
flows, one typically searches for apparent horizons respecting the
physical symmetries. Otherwise, field theory entropy extracted from
the apparent horizon would not obey the symmetries of the state under consideration \cite{Booth:2011qy}. 
This assumption vastly simplifies searches of an apparent horizon in the gravity dual to the boost-invariant flow, as null directions of ingoing and outgoing wavefronts, denoted in the literature by null vectors $l^{a}$ and $n^{a}$, are fixed by symmetry, up to normalization factor, after imposing
\eq
\label{conditionsfornullvectors}
l_{a} l^{a} = 0, \quad n_{a} n^{a} = 0 \quad \mathrm{and} \quad l_{a} n^{a} = -1.
\eqx
The latter condition specifies cross-normalization of the vectors with the minus sign on the right hand side ensuring that $l^a$ is future pointing if $n^{a}$ is and vice versa. In particular, for the metric Ansatz \eqref{e.5d} vectors solving \eqref{conditionsfornullvectors} read
\eqn
l_a &=& h(t,u)\left\{-\frac{a(u)\alpha(t,u)}{\sqrt{2}\sqrt{u}},0,0,0,-\frac{d(t,u)}{2\sqrt{2}u}\right\}, \\
n_a &=& \frac{1}{h(t,u)}\left\{-\frac{a(u)\alpha(t,u)}{\sqrt{2}\sqrt{u}},0,0,0,\frac{\sqrt{2}d(t,u)}{2\sqrt{2}u}\right\},
\eqnx
where $h(t,u)$ is a positive scalar function playing no role in the discussion here.

The condition for an apparent horizon on a given time slice, which is taken here to be $t = const$, is that the rate of change of the area of a lightfront (transverse area of congruence of null geodesics) in $l^{a}$ direction is 0. This quantity is measured by the expansion $\theta_{l}$ given by
\eq
\label{thetal}
\theta_{l} = (g^{a b} + l^{a} n^{b} + l^{b} n^{a}) \nabla_{a} l_{b}
\eqx
with $\theta_{n}$ defined in a completely analogueous way.

In practice, for a given constant time hypersurface in coordinates fixed by the metric Ansatz \eqref{e.5d} with some definite choice of lapse, one searches for $u$'s such that $\theta_{l}$ vanishes and $\theta_{n}$ is negative.
\begin{figure}[t]
\includegraphics[height=6cm]{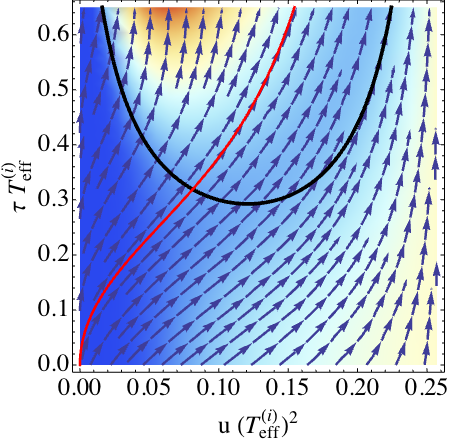}
\caption{The apparent horizon (black curve) and a radial null geodesic 
(red curve) sent from the boundary (left
edge of the plot) at $\tau=0$ into the bulk for a sample profile (no. 23 from table \ref{tab:profs}). This curve
coincides with a curve of fixed Eddington-Finkelstein proper time $\tau_{EF}=0$. Background colors correspond to curvature with blue denoting small curvature and red large curvature. The red region inside the apparent horizon denote the neighborhood of the curvature singularity. One can see (and check the numerical factor explicitly) that curvature at the right edge of the plot remains the same during the evolution, in agreement with expectation that vanishing lapse freezes the time flow there. The curvature at the left edge of the plot also remains constant due to imposed AdS asymptotics.
}
\label{fig:ingoingnullAH}
\end{figure}
Fig. \ref{fig:ingoingnullAH} shows the apparent horizon for a particular choice of lapse and sufficiently large radial cutoff. In all the simulations we did there was no apparent horizon on the initial time slice. However, it was always possible to choose large enough radial cutoff to see it develop later during the simulation. 
Our coordinate frame makes it appear, starting from some time, at two radial positions on each constant time slice, forming a U-shaped structure surrounding the curvature singularity, as shown on Fig. \ref{fig:ingoingnullAH}.

The apparent horizon's entropy density is defined here as Bekenstein-Hawking entropy (density)
\eq
\label{BH.entropy}
s_{AH} = \frac{1}{4 G_{N}} \frac{t}{u} a(u) b(t,u) c(t,u)^2 {\Big |}_{u = u_{AH}}.
\eqx
Left as it is, it is still not a local entropy from the dual field theory point of view. We still need to associate points on the horizon with points at the boundary by providing a so-called bulk-boundary map. Such an association is not free of ambiguities, however here it seems 
that the only sensible way to do so is to associate horizon points with points on the boundary lying on the same ingoing radial null geodesic \cite{Figueras:2009iu,Booth:2009ct}, much in the spirit of Vaidya spacetime and its generalizations \cite{Hubeny:2007xt}. Moreover, due to the large number of symmetries, the direction
of the null geodesic is essentially fixed and unambiguous.
Note that such mapping is consistent with the one introduced in the context of fluid-gravity duality \cite{Bhattacharyya:2008xc}.

For all the considered initial states, we found a non-zero apparent horizon entropy at the boundary time $\tau = 0$, in the sense explained above. Although in the non-equilibrium regime there is no clear microscopic picture of apparent horizon entropy, its initial value compared to the final one is a useful measure of how far from equilibrium a given initial state was. In particular, using that $G_{N} = \f{\pi}{2 N_{c}^{2}}$ \cite{Balasubramanian:1999re} one can reexpress \eqref{BH.entropy} in terms of dimensionless entropy density $s_{n-eq}$ measured in units of the initial effective temperature $T_{eff}^{(i)}$
\eq
\label{eq.dimentropyden}
s_{n-eq} = \frac{s_{AH}}{\frac{1}{2} N_{c}^{2} \pi^2
\big(T_{eff}^{(i)}\big)^{2}},
\eqx
as introduced in section \ref{subsec.bjorkenhydro}.
\section{The numerical simulation\label{sec.numsim}}

\subsection{Determination of $u_0$}

\begin{figure}
\includegraphics[height=6cm]{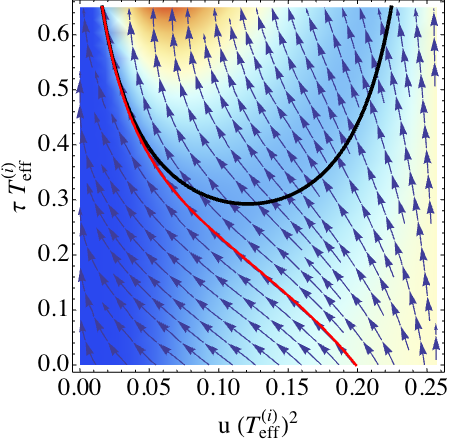}
\caption{Determination of the cutoff point $u_0^{(EH)}$ ensuring long evolution time necessary to see thermalization in the boundary theory for a sample profile (no. 23 from table \ref{tab:profs}). In order to obtain $u_{0}^{(EH)}$, which is expected to approximate well the position of the event horizon at initial time hypersurface, we anticipate that for large enough time the position of the event horizon coincides with the position of the apparent horizon we consider and shoot backwards in time an outgoing null geodesic (almost) tangent to late time apparent horizon (plotted as thick red curve). Other outgoing null geodesics are plotted as arrowed curves. The apparent horizon is plotted as thick black curve. Background colors, as in Fig. \ref{fig:ingoingnullAH} correspond to curvature, from blue (small curvature) to red (large curvature).} 
\label{fig:c1tune}
\end{figure}

As explained in section~\ref{subsec.outerbc}, in order to be able to extend the simulation
to sufficiently large values of proper time and to observe a clean and unambigous
transition to hydrodynamics, it is crucial to optimally tune the position of the
cutoff in the bulk which we denote by $u_0$. The optimal value of $u_0$ would be
the position of the \emph{event} horizon on the initial hypersurface. Then, apart
from purely numerical complications, simulations could be extended indefinitely
and the simulation domain would completely cover the whole exterior of the event horizon, safely separating the evolution from the curvature singularity.

The principal stumbling block is the lack of knowledge of the position of the event horizon
at $t=0$. Moreover, due to the lightlike nature of a part of the initial hypersurface,
we cannot locate an apparent horizon there using the vanishing of expansion scalars.

However, the practical determination of the optimal $u_0$ is in fact quite simple.
First we run our simulation with a relatively large value of $u_0$ for which the
simulation would break down due to the appearance of a curvature singularity 
(as shown in Fig.~1C). An initial guess for this first exploratory
value of $u_0$ is typically provided by taking $u_0$ to be $10-20\%$ larger than the position
of the FG coordinate singularity listed in table \ref{tab:profs}.

The representative outcome for this exploratory simulation is shown in Fig.~\ref{fig:c1tune}. We identify the apparent horizon using the vanishing of the expansion scalar $\theta_l$. To get a first estimate of the position of the event horizon
at $t=0$, we take the outgoing radial null geodesic
from the neighborhood of the late time outer edge of the apparent horizon and evolve it backwards in time until
it reaches the initial time hypersurface. The resulting position $u_0$ will be a very
good estimate of the initial position of the event horizon and will allow for a long
period of evolution.

The initial exploratory simulation was also important for us as we used it to extract
the initial entropy corresponding to the given initial conditions as outlined in
section \eqref{subsec.AH}.

\subsection{The numerical implementation and tests}

In the present work we adopted a Chebyshev discretization of the spatial grid.
This allows us to use quite modest grid sizes ($N=201, 257, 513, 1025$ depending
on the initial profile) and also allows for very accurate evaluation of spatial
derivatives at the AdS boundary which are necessary for extracting the gauge theory
energy-momentum tensor. The downside is that Chebyshev differentiation is very
much nonlocal so any problems at the edges will affect quicky the whole
integration domain. We implemented the Chebyshev differentiation using
Discrete Cosine (Fourier) transforms and the {\tt fftw3} library. 
Time stepping was done using
an adaptive $8^{th}/9^{th}$ order Runge-Kutta method from the GNU scientific 
library\footnote{Recompiled to use {\tt long double} numbers.}.

In order to test the numerical simulations in all cases we monitored the
preservation of the ADM constraints in the form
\eq
\mbox{\rm test}_{C0}\equiv \f{R-K_{ij} K^{ij}+K^2+12}{|R|+|K_{ij} K^{ij}|+K^2+12}
\eqx
and similarly for the momentum constraints.
As an additional check, for some profiles we compared the energies $\eps(\tau)$ 
extracted from simulations done with two different choices of ADM lapse for the same
initial conditions. We found a very good agreement. Another completely independent
cross-check was to compare the $\eps(\tau)$ extracted from the numerical simulation
with $\eps(\tau)$ computed analytically as a power series in $\tau$. 
Within the radius of convergence of the power series, we found excellent
agreement (some example comparisons will be presented
in the following section). Finally, for a fixed profile (no. 23 in Appendix~A), we also checked
whether the dynamical horizon identified for simulations with two different
choices of lapse functions really coincide. This was done by i) comparing the areas
extracted from the intersection of the dynamical horizon with null geodesics 
propagating from the boundary for a range of $\tau$, ii) comparing the curvature
invariants $R_{abcd}R^{abcd}$ at these intersection points.
Again we found perfect agreement.

\section{The profiles\label{sec.profiles}}

In the companion article \cite{Heller:2011ju}, we have presented a thorough analysis 
of the physical characteristics of thermalization of a boost-invariant plasma
system obtained using the numerical setup described in the present paper.

In this section, we would like to discuss in more detail certain aspects which
are interesting in view of the results obtained in \cite{Heller:2011ju}. Firstly,
we discuss various geometrical characteristics of the initial conditions (of which
we give a complete list in Appendix~A) in relation to the event and apparent
horizons, then we describe the qualitative behaviour of the proper time 
evolution of the effective temperature giving evidence that a transient rise
of the tempearture in the non-equilibrium regime observed for initial conditions
with high entropy is a real effect and not a numerical artefact.
Finally, we discuss stability issues of our criterion for thermalization 
(\ref{e.criterion}).

\subsection{Geometric characteristics of the initial states}     

We considered 29 different initial conditions specified by $C_{0}(u)$, as shown in table \ref{tab:profs} in Appendix~A. We focused on profiles, which had no singularity for finite values\footnote{It would be interesting to relax this assumption.} of $u$, 
but we allowed for a possible blow up as 
$u \rightarrow \infty$.  Within the considered class, 
the numerical evolution from initial profiles gave rise to 
quite rich dynamics, exhibiting a wide variety of behaviors. Nevertheless, 
in~\cite{Heller:2011ju} we observed surprising regularities - entropy extracted from
the apparent horizon at $\tau=0$ turns out to characterize the key features
of the transition to hydrodynamics. It is therefore interesting to
understand this in more detail.

As the initial time hypersurface coincides with the one in the Fefferman-Graham coordinates, we can use Fefferman-Graham constraint equation \eqref{constr2} and locate radial position $u_0^{(FG)}$ for which $A_{0}(u)$ vanishes (the coordinate singularity of Fefferman-Graham chart, as originally pointed out in \cite{Beuf:2009cx} and outlined in section \ref{sec.bif}).
\eq
A_{0}(u_0^{(FG)})=0.
\eqx
Although we have no clear geometric interpretation of this particular point, it turns out that there is a surprising correlation between the initial non-equilibrium entropy, as obtained from an apparent horizon using the prescription elucidated in section 
\ref{subsec.AH}, and the radial position of Fefferman-Graham singularity at $t = 0$ (see Fig. \ref{fig:s0_as_a_func_of_uFG}). 

The position of the event horizon on the initial time slice $t = 0$, is to a very good 
degree approximated by the value of $u_{0}$ determined as in 
figure~\ref{fig:c1tune}. We found that it is also correlated with the
position of the Fefferman-Graham singularity (see figure 
\ref{fig:uEH_as_a_func_of_uFG}), and hence with the initial apparent horizon entropy. This result was very helpful in running the actual simulations, 
as it provided a good first guess for the optimal position of the radial cutoff. 

\begin{widetext}

\begin{figure}
\includegraphics[width=15cm]{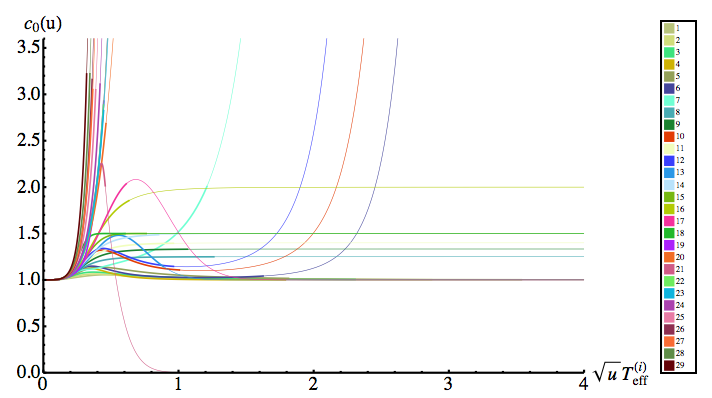}
\caption{Plots of initial warp-factors as functions of square root of radial position. Thicker curves denote parts of initial data outside event horizon on initial time slices. Numbers in the legend match profile numbers collected in 
table \ref{tab:profs} (color online).}
\label{fig:fig_ini_data}
\end{figure} 
\end{widetext}

As outlined in \cite{Heller:2011ju}, it seems that, at least for the class of profiles considered here, the apparent horizon entropy at $\tau = 0$ provides a key infrared characteristic of the initial conditions. One possible explanation, supported by figure \ref{fig:fig_ini_data}, is that the event horizon on the initial time slice effectively removes from the dynamics the part of spacetime for which there is a significant difference between various initial data. 
As a result, the space of initial data on the initial time slice 
\emph{truncated at the event horizon} might be actually quite simple and crudely characterized by a couple of parameters of which the initial non-equilibrium entropy together with the initial effective temperature may be the most important ones. Note also that the initial non-equilibrium entropy is measured on $\tau_{EF} = 0$ Eddington-Finkelstein hypersurface differing from $t = 0$ hypersurfaces 
where our initial conditions are imposed. The close correlation between 
$u_0^{(FG)}$ (on $t=0$) and the initial entropy (defined on $\tau_{EF} = 0$)
is, therefore, quite surprising.
It would be interesting to understand the initial value problem in the ingoing Eddington-Finkelstein coordinates and check whether the mechanism behind the simplicity of initial states conjectured here is indeed correct.

\begin{figure}
\includegraphics[width=9cm]{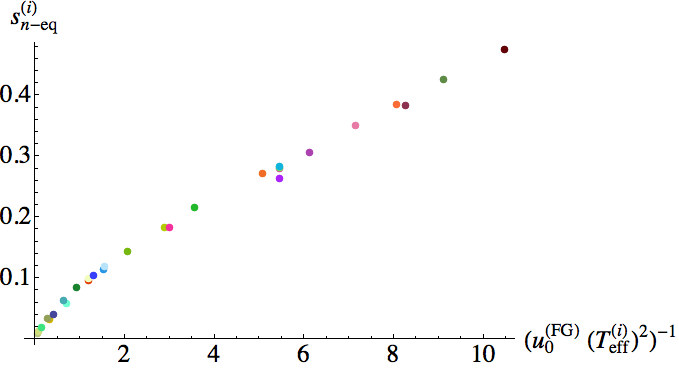}
\caption{Dimensionless initial non-equilibrium entropy given by \eqref{eq.dimentropyden} as a function of the position of Fefferman-Graham singularity showing clear correlation. Both quantities are plotted in the units of the effective temperature at $\tau = 0$. Color code matches figures \ref{fig:fig_ini_data} and \ref{fig:uEH_as_a_func_of_uFG} (color online).}
\label{fig:s0_as_a_func_of_uFG}
\end{figure} 

\begin{figure}
\includegraphics[width=8.5cm]{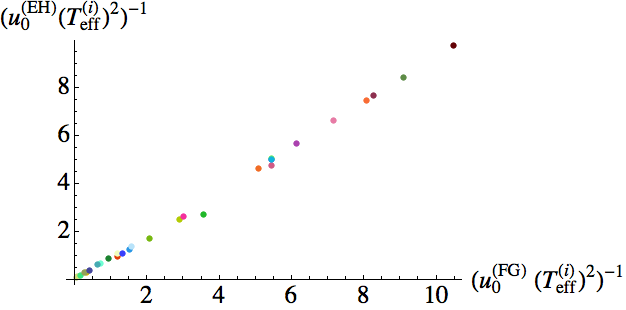}
\caption{Approximate position of the event horizon on the initial time slice as a function of the position of Fefferman-Graham singularity showing clear correlation. Both quantities are plotted in the units of the effective temperature at $\tau = 0$. Color code matches figures \ref{fig:fig_ini_data} and 
\ref{fig:s0_as_a_func_of_uFG} (color online).}
\label{fig:uEH_as_a_func_of_uFG}
\end{figure}  

\subsection{Qualitative features of plasma expansion\\ -- cooling and reheating}

\begin{figure}
\includegraphics[width=8 cm]{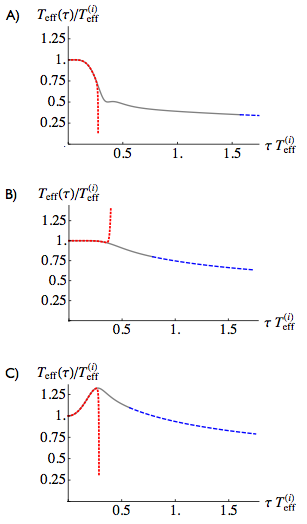}
\caption{Comparison of effective temperatures as functions of time obtained from numerics (solid gray and dashed blue curves) and given by early time power series expression (dotted red curves) for three sample profiles. Situations A), B) and C) correspond respectively to initial profiles 7
, 23 
and 29
from table \ref{tab:profs}, representing initial states with small, intermediate and large initial entropies. Solid gray curves denote far-from-equilibrium part from the evolution, whereas dashed blue curves correspond to hydrodynamic regime with thermalization time set by the criterium \eqref{e.criterion}. Each plot shows perfect agreement between numerical results and early time power series solutions within 
the radia of convergence of the latter. Note also that the radius of convergence of early time power series is much too small to observe the transition to hydrodynamics.}
\label{fig:series}
\end{figure}

The investigations of the transition to hydrodynamics in the companion paper
\cite{Heller:2011ju} revealed that for some initial conditions, the plasma does
not cool but initially undergoes a period of `reheating'. 
Fig. \ref{fig:series} shows the time evolution of the effective temperature for three sample profiles. Depending on the value of initial non-equilibrium entropy, we found three types of behavior. 
Profiles with small initial entropy, as compared to the initial effective temperature, 
led to the effective temperature (and so the energy density) decaying quite rapidly with time. Initial conditions characterized by largest initial entropy among the states we considered, led to a growth of the effective temperature in the initial
stage of evolution followed only later by a cooling phase, as required by Bjorken hydrodynamics. The peculiar feature of the latter states was that the effective temperature at thermalization was for them sometimes higher than the initial one.
This does not mean that thermalization (understood here always as a transition to hydrodynamics) occurred instantaneously but rather that it occurred after a sizable non-equilibrium evolution including a reheating phase.
Finally, we also encountered a profile, with an intermediate value of initial non-equilibrium entropy compared to others we considered, exhibiting initially a plateau in the effective temperature as a function of time, which only later decayed. 

For the three types of the initial states discussed above we compared the energy density obtained from numerical simulation with truncated early time power series obtained directly from the initial profile, as introduced in \cite{Beuf:2009cx} and reviewed in section \ref{subsec.earlytimedynamics}. We observed perfect agreement within the convergence radia of the power series solutions. The agreement constitutes another nontrivial test of our numerics. This also shows that the reheating behavior is 
a real effect and is not due to some pathologies in the numerics.

In all these cases we also see that
the radius of convergence of the power series is smaller than the time of the
transition to hydrodynamics.

\subsection{Stability of the criterion of thermalization\\ -- subtleties for small
entropy initial data}

\begin{figure}
\includegraphics[width=8.5cm]{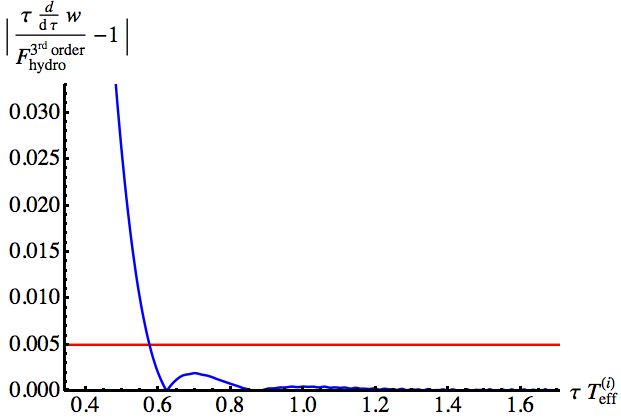}
\caption{Thermalization level (red line) and criterion expression (blue line) plotted as a function of time for the profile no. 29 in table \ref{tab:profs}, with $s^{(i)}_{n-eq}=0.4761$. 
}
\label{fig:crit1}
\end{figure} 

\begin{figure}
\includegraphics[width=8.5cm]{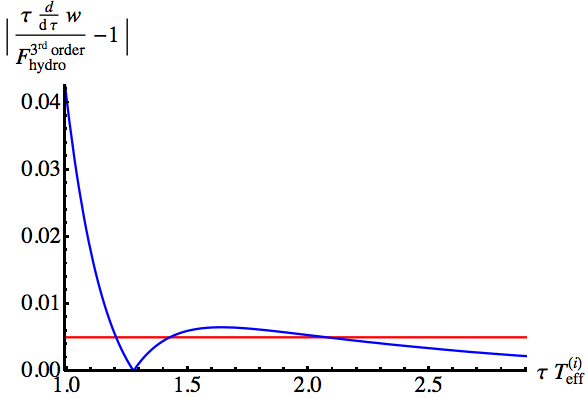}
\caption{Thermalization level (red line) and criterion expression (blue line) plotted as a function of time for the profile no. 3 in table \ref{tab:profs}, with $s^{(i)}_{n-eq}=0.0200$. 
}
\label{fig:crit2}
\end{figure}

Finally, we would like to point out a subtlety in determining the thermalization time, especially relevant for profiles with low initial non-equilibrium entropy. In \cite{Heller:2011ju} we defined thermalization time as the time after which $\f{\tau}{w} \f{d}{d\tau} w$ (with $w$ defined as in \eqref{wdef}) obtained numerically from gravity deviates from third order hydrodynamics result \eqref{e.third} by less than $0.5\%$ (see the criterion \eqref{e.criterion}). As the magnitude of acceptable deviation from hydrodynamics is a free parameter (at least within a reasonable range), thermalization is never a clear cut event. Moreover, the form of $F_{hydro}$ is known only up to the third order in gradients and so it might happen that the thermalization criterium \eqref{e.criterion} is actually sensitive to the inclusion of terms higher order in gradients (fourth order hydrodynamics and higher), whose precise form has not been determined
so far. 

As an example of this subtlety, let us present two figures with the time dependence of the left hand side of our thermalization criterium. Figures \ref{fig:crit1} and \ref{fig:crit2} contain plots of the expression \eqref{e.criterion}, which compares the values of (numerical) exact function $F(w)\equiv \tau \f{d}{d\tau} w$ and the one obtained from third order viscous hydrodynamics $F^{3^{rd} order}_{hydro}(w)$.  
Thermalization time is determined numerically from the \emph{last} intersection 
point between the plot of of \eqref{e.criterion} and the threshold line at 0.005.

In Fig. \ref{fig:crit1} the thermalization time is determined in a very robust way
as within a reasonable range of variation of the threshold, there is just a single point 
of intersection with the threshold line. Moreover, the curve \eqref{e.criterion}
is very steep so modifying the threshold value by a factor of 2 or 4 
($0.01, 0.02$) would lead to only a small change in the determined thermalization
time.

Fig. \ref{fig:crit2} shows an analogous plot for an initial condition
with one of the smallest initial non-equilibrium entropy.
In contrast to Fig. \ref{fig:crit1}, we have three intersection points which
are quite unstable w.r.t. changing the threshold ($0.01$, $0.02$). Indeed, for
these higher threshold values there would be just a single intersection point
located closer to the previous example.
The slowly decaying tail of \eqref{e.criterion} suggests that it may be of
a hydrodynamic nature albeit of a higher order type ($4^{th}$ order and higher).
We expect that including the currently unknown higher order hydrodynamic terms into 
$F_{hydro}(w)$ would bring Fig. \ref{fig:crit2} to look more like 
Fig. \ref{fig:crit1}. Indeed, knowing in principle the all-order $F_{hydro}(w)$,
the function
\eq
\frac{\tau \f{d}{d\tau}w}{F_{hydro}(w)}-1
\eqx
should be exponentially small on the right hand side of the above plots.

We adopted the criterion
with a fixed threshold in order to avoid ambiguities related to subjective judgement which point of intersection to choose.
However, we have to take into account that the small initial entropy data
are much more subtle in this respect. 

\section{Summary and open directions\label{sec.summary}}

This paper introduces an ADM formalism-based scheme for numerical integration of Einstein's equations with negative cosmological constant in the setup describing gravity dual to a boost-invariant strongly coupled plasma system. 
The main motivation for this came from the earlier work by some of us \cite{Beuf:2009cx}, which used the Fefferman-Graham coordinates to constrain the form of initial time ($\tau = 0$) metric components specifying in this way gravity representation of dual far-from-equilibrium initial states (see section \ref{sec.bif}). 
In~\cite{Beuf:2009cx} we also studied time evolution in a power series form,
albeit with a too small radius of convergence to observe directly a transition to
hydrodynamics. Our present numerical approach overcomes this shortcoming.
We are interested, as in~\cite{Beuf:2009cx}, in unforced (i.e. with all sources 
in the gauge theory turned off) relaxation of these states in order to clearly separate 
the thermalization process from the creation of the initial non-equilibrium states 
(as opposed to earlier works \cite{Chesler:2008hg,Chesler:2009cy}).

Numerical treatment of the initial value problem requires specifying it on a compact domain with boundary conditions not altering the evolution, at least outside of the event horizon. Unfortunately, the Fefferman-Graham coordinates used in \cite{Beuf:2009cx} do not provide a sensible notion of bulk cutoff, in particular, based on the example of AdS-Schwarzschild black brane they are not expected to cover interiors of dynamical black branes of interest. Because of this, we introduced a new coordinate frame, which, much in the spirit of the relation between Kruskal-Szekeres and Schwarzschild coordinates in the case of Schwarzschild black hole, coincides with the Fefferman-Graham chart only at the initial time hypersurface, but extends beyond
the Fefferman-Graham coordinate singularity. Moreover, this new
coordinate system allows for a convenient bulk cutoff and, if needed, for 
recovering black brane interior or horizon excision (see section \ref{sec.newcoord} and VD).

The new coordinate frame we introduced allows to adopt ADM formalism-based scheme for numerical evolution (see sections \ref{sec.admreview} and \ref{sec.admforbif}). The key technical element of the present work, which to our knowledge has not been explored in the literature before, is a new treatment of the outer boundary condition
in a potentially highly curved spacetime through forcing the lapse 
to vanish at a fixed radial position. This stops the flow of time at a given point (actually on a codimension-2 hypersurface) leading to a very convenient bulk cutoff for numerical integration, as we are free to specify its position. 

In order to ensure a long enough simulation time to see the transition to a hydrodynamic regime in the dual gauge 
theory, such a cutoff should be very close to the position of the \emph{event} 
horizon on the initial time slice. Although the latter information is not available from the start due to the teleological nature of the event horizon, we nevertheless managed to find a simple way of determining it. To do so, we ran a trial simulation with large enough bulk cutoff to eventually see an apparent horizon forming and then traced back outgoing radial null geodesic tangent to the late time apparent horizon. As in the late time regime we expect our apparent horizon to coincide with the event horizon, the trajectory of outgoing null geodesic of interest should,
to a very good degree, coincide with the location of the event horizon. The intersection point of this trajectory with the initial time hypersurface provides the locus 
where the lapse needs to vanish in order to ensure a long simulation time.

Another nontrivial issue which we encountered, is connected with
imposing asymptotically AdS boundary conditions\footnote{This was an important 
point, as we wanted to ensure that the plasma system we are studying evolves in ordinary
Minkowski spacetime and not in some time-dependent background geometry.}, 
as well as obtaining the expectation value of the gauge theory energy-momentum tensor 
carrying crucial information 
about the thermalization process (see subsections \ref{subsec.bdrycondAdSbdry} and \ref{subset.formulasfordualstresstensor}).  
The chief difficulty came from the fact that due to the dynamical time-dependent
radial metric coefficient, the points on constant radius hypersurfaces approach 
the asymptotic boundary at various rates.
This complicates both obtaining the form of the boundary metric, as well as 
invalidates the ``standard'' formula for dual stress tensor expressed 
in terms of extrinsic curvature of constant radius hypersurface. 
To circumnavigate this issue we performed near-boundary coordinate 
transformation to the Fefferman-Graham coordinates, in which it is 
particularly simple to obtain expressions for the boundary metric 
and expectation value of the gauge theory energy-momentum tensor.

One interesting finding of our studies is that, because of fixing bulk gauge 
freedom by specifying the form of a time-like warp-factor, the boundary and 
bulk notions of time differed by a boundary diffeomorphism. This also altered 
the form of the Minkowski metric on the boundary. We expect that similar issues 
might arise when trying to impose normalizable boundary conditions when 
studying e.g. spectrum of quasinormal modes in spacetimes expressed in coordinate 
frames with the gauge fixed by specifying the form of warp-factors associated with the field theory directions.

The accuracy of our numerical simulation, based on spectral discretization in radial direction and finite difference time stepping using high order Runge-Kutta method, has been monitored in a couple of ways (see section \ref{sec.numsim}). In the first place, we used two appropriately normalized constraint equations, which, if small at all points of the grid, provide a nontrivial check of the numerics, as we used an unconstrained evolution scheme. The other way of making sure that our results are correct, was to run the numerics for a given initial profile with different choices of lapse (both the position at which the lapse vanishes and its functional 
dependence on the dynamical metric components). As various choices of lapse cover various parts of underlying manifold foliating it in various ways by constant time hypersurfaces, obtaining for each choice the same (up to numerical error) effective temperature as well as apparent horizon entropy
as functions of boundary time gave highly nontrivial indications of the accuracy of numerical simulation. The third way of making sure the numerics worked fine was to compare the effective temperature as a function of time obtained numerically with its analytic form in the early time power series (see section \ref{sec.profiles}B). Again, within the radius of convergence of the early time power series we observed perfect agreement.

A very surprising finding of this work, reported first in the companion article \cite{Heller:2011ju}, is that characteristics of thermalization for all the profiles we considered are correlated with the initial value of non-equilibrium entropy defined on the apparent horizon (see subsection \ref{subsec.AH} for details on what we mean here by local non-equilibrium entropy). A possible explanation of this might be that the event horizon appears on the initial time hypersurface as soon as the initial time warp-factors start to differ significantly from their vacuum value. It would be very interesting to verify this finding by considering initial conditions in the Eddington-Finkelstein frame, as in this case, due to constant time foliation being aligned along radial ingoing null geodesics, the apparent horizon would possibly appear from the start (see however \cite{Heller:2012km}). It would be thus interesting to understand the correlation between the position of the apparent horizon (if it exists) on the initial time hypersurface with the level of complication of a warp-factor setting initial conditions. It would also be interesting to perform explicit numerical simulations and check how the position of the event horizon is correlated with the position of the apparent horizon on the initial time hypersurfaces in the Eddington-Finkelstein coordinates.

Another interesting point is related to the choice of initial profiles we considered here. We focus only on the profiles which are either regular for all radii or blow up as the radial coordinate is taken to infinity. In principle, it is possible that there are initial profiles diverging for a finite radial position, but having the event horizon shielding the singularity. It would be interesting to investigate this issue in more detail, as such studies might uncover a new class of initial states.

The profiles we considered exhibit a wide variety of behaviors in the bulk. Although, as expected on general grounds, the evolution of the one-point function of the stress tensor do not depend on the details of a given profile hidden behind the event horizon, nonlocal observables such as entanglement entropy, Wilson loops or higher point correlation functions generally do \cite{AbajoArrastia:2010yt}. It would be thus interesting to investigate this issue in our setup, as has been done for the Vaidya spacetime in \cite{Balasubramanian:2010ce,Balasubramanian:2011ur}. Our setup might be advantageous compared to setting initial conditions in Eddington-Finkelstein type of coordinates when it comes to obtaining equal time two-point functions of heavy scalar operators, as these are approximated by spacelike geodesics, which might closely align along our initial constant time slices.

Yet another interesting direction to follow would be to perform linear stability analysis of the early time boost-invariant plasma with respect to perturbations breaking boost-invariance (i.e. depending on rapidity), as such instabilities have been found in the weak coupling regime within the colour glass condensate approach 
\cite{Romatschke:2005pm}. 

One direction, which has not been explored at all so far, is to understand holographic thermalization in the gravity backgrounds dual to non-conformal strongly coupled gauge theories. It would be of obvious interest to study in such gauge theories boost-invariant flow, as the
simplest phenomenologically interesting model of plasma dynamics with the use of holographic methods introduced here or in \cite{Chesler:2009cy}.

Finally, it would be fascinating to introduce transverse dynamics in the holographic model of Bjorken flow to get a handle on thermalization phase of the elliptic flow \cite{Snellings:2011sz}, as well as understand how nontrivial structures in the transverse plane at the initial time affect the initial conditions for hydrodynamic evolution \cite{Muller:2011bb}. One motivation for doing this, stemming from the results of this work reported in \cite{Heller:2011ju} (and also earlier in \cite{Chesler:2009cy}), is that in the one-dimensional boost-invariant flow non-Abelian plasma can be very anisotropic, yet well-described by hydrodynamics. Allowing the plasma system to expand in traverse directions should lower, at least intuitively, the degree of anisotropy at thermalization. 
Investigating this issue in more detail is important, as this touches upon 
the question whether the QCD quark-gluon-plasma produced in heavy-ion
collisions indeed needs to be approximately isotropic before the transition to
hydrodynamics.  

\smallskip

\noindent{\bf Acknowledgments:} We thank Andrzej Rostworowski for collaboration 
during the initial stages of the project, Wojciech Florkowski, 
Jean-Yves Ollitrault and Tomasz Roma{\'n}czukiewicz for answering many questions,
Jean-Paul Blaizot for an interesting comment on anisotropy.
This work was partially supported 
by Polish science funds as research projects N N202 105136 (2009-2012) and 
N N202 173539 (2010-2012), as well as the Foundation for Fundamental Research on Matter (FOM) being a part of the Netherlands Organisation for Scientific Research (NWO).
MH and RJ acknowledge the support and hospitality of the Kavli Institute for
Theoretical Physics. This research was supported in part by the National Science Foundation under Grant No. NSF PHY11-25915. 

\appendix
\onecolumngrid
\section{A summary of the initial conditions and simulation results}

\noindent{TABLE I:} Below we collect information about the initial profiles we considered: $C_{0}(u)$ is the initial condition as discussed in section \ref{sec.bif} with $\gamma = \frac{1}{2}\sqrt{3}\pi^{2} \big(T_{eff}^{(i)}\big)^2$; $u_{0}^{(FG)}$ is a position of corresponding coordinate singularity in the Fefferman-Graham chart; $u_{0}^{(EH)}$ is the approximate position of the event horizon on the initial time slice; $w^{(th)}$ is thermalization time measured in terms of an effective temperature at thermalization as defined by equation \eqref{wdef}; $\tau^{(th)} T_{eff}^{(i)}$ is the thermalization time in units of the initial effective temperature; $T_{eff}^{(th)} / T_{eff}^{(i)}$ is the ratio of the effective temperature at thermalization to the initial effective temperature; $s_{n-eq}^{(i)}$ is dimensionless initial entropy density \eqref{eq.dimentropyden} defined by apparent horizon, as described in section \ref{subsec.AH}; $s_{n-eq}^{(f)}$ is the final entropy density obtained from perfect fluid hydrodynamics.

\begin{longtable}{| c | c | c | c | c | c | c | c | c |}
\hline
 \text{No.} & $C_0(u)$ & $u_0^{(FG)}\big(T_{eff}^{(i)}\big)^{2}$ & $u_0^{(\text{EH})}\big(T_{eff}^{(i)}\big)^{2}$ & $w^{(th)}$ & $\tau^{(th)}T_{eff}^{(i)}$ & $T_{eff}^{(th)}/T_{eff}^{(i)}$ & $s_{n-eq}^{(i)}$ & $s_{n-eq}^{(f)}$ \\ \hline
 1 & $\frac{\left(1-\frac{1}{\gamma  u+1}\right) \tanh \left(\frac{\gamma  u}{2}\right)}{4 \gamma  u+1}+1$ & 23.9980
   & 12.5030 & 0.4853 & 3.2287 & 0.1503 & 0.0086 & 0.0120 \\ \hline
 2 & $\frac{1-\frac{1}{\gamma ^2 u^2+1}}{2 (3 \gamma  u+1)}+1$ & 12.8410 & 8.2419 & 0.4868 & 2.6686 & 0.1824 & 0.0127
   & 0.0178 \\ \hline
 3 & $\frac{\gamma  u \left(1-\frac{1}{\gamma  u+1}\right)}{2 \left(2 \gamma ^2 u^2+1\right)}+1$ & 7.0707 & 5.3189 &
   0.4722 & 2.0759 & 0.2275 & 0.0200 & 0.0270 \\ \hline
 4 & $\frac{\gamma ^2 u^2}{2 \left(\gamma ^2 u^2+1\right)^2}+1$ & 3.0462 & 3.2053 & 0.5209 & 1.8717 & 0.2783 & 0.0322
   & 0.0435 \\ \hline
 5 & $\frac{\gamma  u \left(1-\frac{1}{\gamma  u+1}\right)}{2 \left(\gamma ^2 u^2+1\right)}+1$ & 3.6404 & 3.2898 &
   0.4153 & 2.0940 & 0.2963 & 0.0333 & 0.0420 \\ \hline
 6 & $\frac{\gamma ^2 u^2 e^{\frac{\gamma  u}{6}} \left(1-\gamma ^4 u^4 e^{-6 \gamma ^2 u^2}\right)}{2 \left(\gamma
   ^2 u^2+1\right)^2}+1$ & 2.4453 & 2.6413 & 0.3713 & 1.4999 & 0.3433 & 0.0402 & 0.0522 \\ \hline
 7 & $\frac{\gamma ^2 u^2 e^{\frac{\gamma  u}{6}} \left(\frac{\gamma ^6 u^6}{8}+\frac{\gamma ^4
   u^4}{4}+\frac{1}{2}\right)}{\left(\gamma ^4 u^4+\gamma ^2 u^2+1\right)^2}+1$ & 1.4390 & 1.4639 & 0.5507 & 1.5759
   & 0.3494 & 0.0592 & 0.0713 \\ \hline
 8 & $\frac{\gamma ^2 u^2}{2 \left(2 \gamma ^2 u^2+1\right)}+1$ & 1.6148 & 1.5908 & 0.5200 & 1.4265 & 0.3646 & 0.0641
   & 0.0923 \\ \hline
 9 & $\frac{\gamma ^2 u^2}{2 \left(\frac{3 \gamma ^2 u^2}{2}+1\right)}+1$ & 1.0747 & 1.1372 & 0.5625 & 1.3841 &
   0.4064 & 0.0847 & 0.1217 \\ \hline
 10 & $\frac{\gamma ^2 u^2 e^{\frac{\gamma  u}{6}}}{2 \left(\frac{\gamma ^2 u^2}{2}+1\right)^2}+1$ & 0.8425 & 1.0125
   & 0.5634 & 1.2806 & 0.4400 & 0.0968 & 0.1152 \\ \hline
 11 & $\frac{\gamma ^2 u^2}{2 \left(\frac{5 \gamma ^2 u^2}{4}+1\right)}+1$ & 0.8437 & 0.9252 & 0.5809 & 1.3300 &
   0.4368 & 0.0998 & 0.1438 \\ \hline
 12 & $\frac{\gamma ^2 u^2 e^{\frac{\gamma  u}{5}}}{2 \left(\frac{\gamma ^2 u^2}{2}+1\right)^2}+1$ & 0.7676 & 0.9264
   & 0.5768 & 1.2787 & 0.4511 & 0.1039 & 0.1231 \\ \hline
 13 & $\frac{1}{2} \gamma ^2 u^2 e^{-\frac{3 \gamma  u}{4}}+1$ & 0.6609 & 0.7884 & 0.6691 & 1.5080 & 0.4437 & 0.1139
   & 0.1658 \\ \hline
 14 & $\frac{\gamma ^2 u^2}{2 \left(\gamma ^2 u^2+1\right)}+1$ & 0.6430 & 0.7269 & 0.5978 & 1.2556 & 0.4761 & 0.1199
   & 0.1748 \\ \hline
 15 & $\frac{1}{2} \tanh ^2\left(\frac{\gamma ^2 u^2}{25}+\gamma  u\right)+1$ & 0.4884 & 0.5788 & 0.6154 & 1.1825 &
   0.5204 & 0.1441 & 0.2131 \\ \hline
 16 & $\frac{\gamma ^2 u^2}{2 \left(\frac{\gamma ^2 u^2}{2}+1\right)}+1$ & 0.3475 & 0.4006 & 0.5098 & 0.7948 & 0.6414
   & 0.1826 & 0.2825 \\ \hline
 17 & $\frac{1}{2} \gamma ^2 u^2 e^{-\frac{\gamma  u}{2}}+1$ & 0.3336 & 0.3764 & 0.5396 & 0.8448 & 0.6388 & 0.1841 &
   0.2919 \\ \hline
 18 & $\frac{1}{2} \tanh ^2\left(\gamma ^2 u^2+\gamma  u\right)+1$ & 0.2807 & 0.3652 & 0.6139 & 0.9465 & 0.6486 &
   0.2168 & 0.3307 \\ \hline
 19 & \pbox{20cm}{$\exp{\left(\frac{1}{2}\int_{0}^{\gamma u} \left\{\sqrt{2 v_{+}'(x)-v_{+}(x)^2} - v_{+}(x)\right\}dx\right)}$\\ with $v_{+}(x) = \tan{x}-\tanh{(x+x^{4})}$} & 0.1838 & 0.1996 & 0.6150 & 0.7719 & 0.7968 & 0.2642 & 0.5009 \\ \hline
 20 & $\frac{\gamma^2 u^2}{2}+1$ & 0.1971 & 0.2147 & 0.6111 & 0.7817 & 0.7817 & 0.2711 & 0.4797 \\ \hline
 21 & \pbox{20cm}{$\exp{\left(\frac{1}{2}\int_{0}^{\gamma u} \left\{\sqrt{2 v_{+}'(x)-v_{+}(x)^2} - v_{+}(x)\right\}dx\right)}$\\ with $v_{+}(x) = \tan{x}-\tanh{(x-\frac{1}{4}x^{4})}$} & 0.1838 & 0.2097 & 0.6548 & 0.8525 & 0.7681 & 0.2803 & 0.4891 \\ \hline
 22 & \pbox{20cm}{$\exp{\left(\frac{1}{2}\int_{0}^{\gamma u} \left\{\sqrt{2 v_{+}'(x)-v_{+}(x)^2} - v_{+}(x)\right\}dx\right)}$\\ with $v_{+}(x) = \tan{x}-\tanh{(x+\frac{1}{6}x^{4})}$} & 0.1838 & 0.1979 & 0.6346 & 0.7952 & 0.7980 & 0.2810 & 0.5146 \\ \hline
 23 & $\cosh (\gamma  u)$ & 0.1838 & 0.1987 & 0.6306 & 0.7886 & 0.7997 & 0.2839 & 0.5142 \\ \hline
 24 & $e^{\frac{\gamma ^2 u^2}{2}}$ & 0.1634 & 0.1762 & 0.6453 & 0.7682 & 0.8401 & 0.3062 & 0.5778 \\ \hline
 25 & $\cosh \left(\frac{3 \gamma ^2 u^2}{10}+\gamma  u\right)$ & 0.1398 & 0.1503 & 0.6380 & 0.7029 & 0.9077 & 0.3501
   & 0.6687 \\ \hline
 26 & $\gamma ^4 u^4+\frac{\gamma ^2 u^2}{2}+1$ & 0.1210 & 0.1303 & 0.6293 & 0.6460 & 0.9742 & 0.3838 & 0.7624 \\ \hline
 27 & $\frac{1}{2} \gamma ^2 u^2 e^{\gamma  u}+1$ & 0.1243 & 0.1338 & 0.6324 & 0.6571 & 0.9624 & 0.3859 & 0.7465 \\ \hline
 28 & $\cosh \left(\frac{4 \gamma ^2 u^2}{5}+\gamma  u\right)$ & 0.1099 & 0.1187 & 0.6356 & 0.6227 & 1.0207 & 0.4259
   & 0.8433 \\ \hline
 29 & $\frac{1}{2} \gamma ^2 u^2 e^{2 \gamma  u}+1$ & 0.0955 & 0.1026 & 0.6298 & 0.5754 & 1.0947 & 0.4761 & 0.9634 \label{tab:profs}
 \\ \hline
 \end{longtable}

\section{Full Einstein equations}

\label{app:b}

The following are Einstein equations for all ADM functions. Due to the size of expressions for $L,M,P$, they were exported from a notebook and follow standard {\it Mathematica} notation for partial derivatives.

\eqn
\frac{\partial b}{\partial t}&=&\frac{-b^2+\alpha L}{t b} \\
\frac{\partial c}{\partial t}&=&\frac{a\alpha M}{c} \nonumber\\
\frac{\partial d}{\partial t}&=&\frac{a\alpha P}{d} \nonumber
\eqnx

\begin{eqnarray}
\label{eq:dtL}
\frac{\partial L}{\partial t} &=& 
-\frac{4 t  u a a'' \alpha  b^2}{d^2}-\frac{4 t  u a a' \alpha_u b^2}{d^2}-\frac{12t  u a a' \alpha  b b_u}{d^2}-  
 \frac{8 t  u a a' \alpha  b^2 c_u}{c d^2}+\frac{4 t  u a a' \alpha  b^2 d_u}{d^3}+\frac{8 t  a a' \alpha  b^2}{d^2}-\frac{4 t  u a'^2 \alpha  b^2}{d^2}- \nonumber \\
&& \frac{4 t  u a^2 \alpha_u b b_u}{d^2}+\frac{2 t  a^2 \alpha_u b^2}{d^2}-  
 \frac{8 t  u a^2 \alpha  bb_u c_u}{c d^2}+\frac{4 t  u a^2 \alpha  b b_u d_u}{d^3}+\frac{6 t  a^2 \alpha  b b_u}{d^2}- \frac{4 t  u a^2 \alpha  b b_{uu}}{d^2}+\frac{4 t  a^2 \alpha  b^2 c_u}{c d^2}- \nonumber\\ 
&&   
 \frac{2 t  a^2 \alpha  b^2 d_u}{d^3}-\frac{4 t  a^2 \alpha  b^2}{u d^2}+\frac{4 t  a^2 \alpha  b^2}{u}- 
 \frac{2 a \alpha  L M}{c^2}-\frac{a \alpha  L P}{d^2}-\frac{\alpha  L^2}{t  b^2}-\frac{L}{t }  \\
\label{eq:dtM}
\frac{\partial M}{\partial t} &=& 
 \frac{8 u a' \alpha  c_u c}{d^2}-\frac{4 a' \alpha  c^2}{d^2}+\frac{4 u a \alpha_u c_u c}{d^2}-  
\frac{2 a \alpha_u c^2}{d^2}+ \frac{4 u a \alpha b_u c_u c}{b d^2}-\frac{2 a \alpha  b_u c^2}{b d^2}- \frac{4 u a \alpha  c_u c d_u}{d^3}- \frac{8 a \alpha  c_u c}{d^2}+ \nonumber \\
&& \frac{4 u a \alpha  c_{uu} c}{d^2}+ \frac{4 u a \alpha  c_u^2}{d^2}+ \frac{2 a \alpha  c^2 d_u}{d^3}+ 
 \frac{4 a \alpha  c^2}{u d^2}-\frac{4 a\alpha  c^2}{u}-  
\frac{a \alpha  M P}{d^2}+\frac{\alpha  L M}{t  b^2} \\
\frac{\partial P}{\partial t} &=& 
8 u a' \alpha_u+\alpha  \left(\frac{8 u a' b_u}{b}-\frac{8 u a' d_u}{d}+\frac{L P}{t  b^2}+8 u a''\right)+  
 4 u a \left(\alpha_{uu}-\frac{\alpha_ud_u}{d}\right)+a\alpha  \left(\frac{4 d_u \left(-\frac{u b_u}{b}-\frac{2 u c_u}{c}+2\right)}{d}+\right.  \nonumber \\
 && \left. \frac{4 u b_{uu}}{b}+\frac{8 u c_{uu}}{c}-\frac{2 M P}{c^2}+\frac{P^2}{d^2}-\frac{4 d^2}{u}+\frac{4}{u}\right) 
\end{eqnarray}

Equation for the evolution of lapse function $\alpha(t,u)$ was obtained from the algebraic expression specifying it by differentiation with respect to time. 

\twocolumngrid

\end{document}